\newcommand*\dsk{./}
\newcommand*\figspath{\dsk/figs}

\documentclass[fontsize=12pt]{article}

\usepackage{amsmath}
\usepackage{amssymb}

\usepackage{subcaption}
\usepackage{lmodern,sansmath}
\usepackage{textcomp}

\usepackage{microtype}

\usepackage{natbib}
\usepackage{graphicx}
\usepackage{booktabs}
\usepackage{upgreek}
\usepackage{color}
\usepackage{fixmath}

\usepackage[colorlinks=true, linkcolor=blue, citecolor=blue]{hyperref}         

\usepackage{xcolor}
\usepackage{soul}
\usepackage{textcase}
\usepackage{bm}
\usepackage{algorithm}
\usepackage{algorithmic}
\usepackage{amsmath}
\usepackage{amssymb}
\usepackage{fancyhdr,lastpage}
\usepackage{enumerate}
\usepackage{graphicx}
\usepackage{epstopdf}
\usepackage{overpic}
\usepackage{caption}
\usepackage{subcaption}

\newcommand{\moy}[1]{\left\langle {#1} \right\rangle}

\def\prodvec{\times}

\definecolor{myred}{rgb}{0.8,0.1,0.1}

\def\Blue#1{{\color{blue}{#1}}}

\def\ds{\displaystyle}

\title{Higher-order statistics and intermittency of a two-fluid HVBK quantum turbulent flow\\  \Blue{to appear in Journal of Fluid Mechanics, 2023.}}

\author{Z. \ Zhang$^1$, I. \ Danaila$^1$, E. \ L\'ev\^eque$^2$, L. \ Danaila$^3$} 
\date{\small $^1$ Universit{\'e} de Rouen Normandie, Laboratoire de Math{\'e}matiques Rapha{\"e}l Salem,  CNRS UMR 6085, Avenue de l'Universit{\'e},  76801 Saint-{\'E}tienne-du-Rouvray, France \\
	\small  $^2$Laboratoire de M{\'e}canique des Fluides et d'Acoustique,
	{\'E}cole Centrale de Lyon,  CNRS UMR 5509, 36 avenue Guy de Collongue, Lyon, France \\
	\small  $^3$Universit{\'e} de Rouen Normandie, Laboratoire Morphodynamique Continentale et C{\^o}ti{\`e}re,  CNRS UMR 6143, Place Emile Blondel, 76821 Mont Saint Aignan, France}

\begin{document}
\maketitle



\begin{abstract}
	
	The Hall-Vinen-Bekharevich-Khalatnikov (HVBK) model is widely used to numerically study quantum turbulence in superfluid helium. Based on the two-fluid model of Tisza and Landau, the HVBK model describes the normal (viscous) and superfluid (inviscid) components of the flow using two Navier-Stokes type equations, coupled through a mutual friction force term. This feature makes the HVBK model very appealing in applying statistical tools used in classical turbulence to study properties of quantum turbulence. A large body of literature used  low-order statistics (spectra, or second-order structure functions in real space) to unravel exchanges between the two fluids at several levels. The novelty in this study is to use a theoretical approach based on first principles to derive transport equations for the third-order moments for each component of velocity. New equations involve the fourth-order moments, which are classical probes for internal intermittency at any scale, revealing the probability of rare and strong fluctuations.
    Budget equations are assessed through Direct Numerical Simulations (DNS) of the HVBK flow based on accurate pseudo-spectral methods. We simulate a forced homogeneous isotropic turbulent flow with Reynolds number of the  normal fluid (based on Taylor's microscale) close to 100.  Values from 0.1 to 10 are considered for the ratio between the normal and superfluid densities. For these flows, an inertial range is not discernible and the Restricted Scaling Range (RSR) approach is used to take into account the Finite Reynolds Number (FRN) effect.   We analyse the importance of each term in budget equations and emphasize their role in energy exchange between normal and superfluid components. Some interesting features are observed: i) transport and pressure-related terms are dominant, similarly to single-fluid turbulence; ii) the mathematical signature of the FRN effect is weak  in the transport of the third-order moment, despite the low value of the Reynolds number;   iii)  for the normal fluid at very low temperatures, the mutual friction  annihilates the effects of viscosity within the RSR. 
   The flatness of the velocity derivatives is finally studied through the transport equations and their limit for very small scales, and it is shown to  gradually increase  for lower and lower temperatures, for both the normal fluid and the superfluid. This similarity highlights the strong locking of the two fluids. The flatness factors are also found in reasonable agreement with classical turbulence. 
\end{abstract}


\pagebreak

\section{Introduction}

Liquid helium below the critical (lambda) temperature $T_\lambda =2.17K$ is a quantum fluid, also called He II. Following the two-fluid concept suggested by  \cite{Tisza_1938} and reformulated and enriched by   \cite{Landau1941}, He II is represented as a  mixture of two fluids with independent velocity fields: a {\em normal} viscous fluid and an inviscid {\em superfluid}. A detailed recount of the historical events leading to the two-fluid model is offered by \cite{QT-Balibar-Tisza}.  A striking feature of the superfluid component is the nucleation of quantized vortices, with fixed (quantized) circulation  and fixed core diameter (of the atomic size). Stretching or viscous diffusion of vortices, which are essential vortex phenomena in classical fluids, are absent in the superfluid component. Complex interactions between quantized vortices lead to Quantum turbulence (QT), a relatively young investigation field opened by Vinen's 1957 experiments on thermally induced counterflow in He II  (see the review by \cite{QT-review-2002-vinen}). Since then, considerable experimental and theoretical efforts (see dedicated reviews or volumes by \cite{QT-book-2009-tsubota,QT-book-sreeni-2012,Barenghi_2013,Barenghi_etal_2014}) were devoted to unravel properties of QT and underline similarities or differences with Classical turbulence (CT). 

Several investigation paths were explored for the study of QT. Since it is admitted that in He II below 0.3K, the normal fluid fraction is negligible, important focus was given to  characterize QT in the superfluid flow.  This state is also referred as superfluid turbulence, or {\em vortex tangle turbulence}, since it is generated in an inviscid flow from the interaction of a large number of quantized vortices tangled in space. Quantized vortices being topological line defects, with infinite velocity and singular vorticity at the centreline, they can be modelled by  'Vortex filament' methods. In such methods, the vorticity is represented by Dirac distributions localised at vortex line locations, which are moved following the Biot-Savart-Laplace law  for the velocity induced by neighbouring lines. Phenomenological models for vortex reconnection are applied. Since the pioneering work by Schwarz  in 1980s, numerous numerical studies of superfluid turbulence using the 'Vortex filament'  method were published (see the recent review by \cite{QT-review-2017-tsubota-num} and citations therein). Another model used for inviscid superfluid turbulence was  the Gross-Pitaevskii equation, which is a  nonlinear Schr{\"o}dinger  equation describing at macroscopic level a quantum system of weakly interacting bosons, as in Bose-Einstein condensates. 
Even though the GP model offers only a partial description of the complexity of superfluid helium, it was extensively used  to explore properties of superfluid turbulence in an ideal setting containing only the superfluid \citep{Nore97a,Abid2003509,dan-2021-CPC-QUTE}.
 
 Considering simultaneously the viscous and inviscid components of He II in a global model is a difficult problem, since characteristic scales range from Angstrom (size of the quantized vortex) to meter (size of the container). The Hall--Vinen--Bekharevich--Khalatnikov (HVBK)  model \citep{Hall_Vinen_1956,QT-book-1965-khal,2009_Donnelly_yearophysic} follows the original idea of the two-fluid model. The  Navier-Stokes (NS) model describes the normal fluid motion and the superfluid motion is defined by an Euler-like equation \citep{Roberts_1974}.  The two fluids do not slip one over other, as they are coupled through a friction force. The improvement over the original two-fluid model is that the expression of the friction force takes into account the influence of quantized vortices through a coarse-grained averaged superfluid vorticity. The average is considered  over an ensemble of parallel (polarized) vortex filaments and uses Feynman's rule to find an equivalent solid-body vorticity for a dense vortex bundle of line density $\mathcal{L}$. Derived initially for two-dimensional or rotating QT, the HVBK was widely used to study QT for general settings. 
 
  Recent modelling efforts were focused on more realistic estimations of the vortex line density
  using approaches considering $\mathcal{L}$ as an independent variable, described by an additional evolution equation (based essentially on Vinen's equation) \citep{QT-book-1991-donnelly,QT-Lipniacki-2006,QT-review-2013-nemirovskii,nemirovskii2020}.   
  
  In \cite{MONGIOVI2018}, the averaged vortex line density per unit volume  was introduced and its evolution equations were considered, for homogeneous, inhomogeneous, isotropic and anisotropic situations. 
  \cite{jou2011hydrodynamic} studied   the effects of anisotropy and polarization in the hydrodynamics of inhomogeneous vortex tangles, thus generalizing the  HVBK equations. These effects contribute to the mutual friction force between normal and superfluid components and to the vortex tension force. An additional equation for the vortex line density was proposed. Applications pertained to  rotating counterflows, flow behind a cylinder, and other types of superfluid turbulence. 
  
  Other recent contributions \citep{QT-coupling-2018-tsu,QT-coupling-2020-gal}  use ideas from the HVBK expression of the friction force to derive models for coupling NS equations with vortex filaments dynamics for superfluid vortices. These NS-VF models, which also include phenomenological approximations, are not discussed in this contribution.  
  
  These models are still flow dependent and a general theory of coupling Navier-Stokes equations with quantized vortex effects is not yet available \citep{nemirovskii2020}. 
   
 The focus of this paper is the detailed investigation of turbulent dynamics of the HVBK model, considered in its original form. The HVBK model has the merit to 
 provide a physically consistent closed set of equations for the coarse-grained (two-fluid) dynamics of He II, and to yield results in  agreement with 
 experimental studies of He II  \citep{Roche_etal_2009,Salort_etal_2010,Salort_etal_2012,Baggaley_2012,Boue_2015,Biferale_2018}. The analysis presented here is based on Direct Numerical Simulations (DNS) of the model and thus could be easily adapted to further evolutions of the HVBK or other equivalent QT models based on Navier-Stokes type equations.  
 We adapt statistic analysis tools originally developed for CT governed by classical Navier-Stokes equations. 
 Exploring similarities between CT and QT has been a permanent guideline for studying QT, \citep{QT-review-sreeni-2012,QT-book-sreeni-2012}. 
 
 The novelty of this study  is to push the analysis to high-order moments of each component of velocity, with the aim to probe internal intermittency, i.e. assess the 4th-order structure function, and the corresponding flatness of the velocity derivative. Previous contributions used low-order statistics (spectra, or second-order structure functions in real space) to describe exchanges between the two fluids. We derive transport equations for the 3rd-order moments based on first principles. New equations involve the 4th-order moments, which are classical probes for internal intermittency at any scale.  The general purpose of this contribution is therefore to build new bridges between CT and QT, as explained in detail below.
 
Previous studies have noted that QT in He II has a lot in common with CT. 
Experimental studies focused on the total velocity of the fluid, are unable, as yet, to distinguish between the normal and the superfluid components.  Several authors \citep{Maurer_1998, Roche_2007, Bradley_2008, Salort2010Specmeasure, Salort_etal_2012} have reported that, in the inertial range, the isotropic and homogeneous quantum turbulence velocity spectrum has a $(-5/3)$ scaling law. The effective spectrum of superfluid vorticity (superfluid vortices averaged on a volume much larger than the inter-vortex length scale) scales as $1/3$. Scaling laws such as $5/3$, or $1/3$ for the vorticity, are predicted by Kolmogorov theory and are well established for classical turbulence, when the Reynolds number of the flow is large enough \citep{Djenidi2017PRF}. 
	Numerical studies of QT have proved the same large-scale behavior using the HVBK, 'Vortex filament' or GP models (see recent review by  \cite{QT-review-2017-tsubota-num}).

	Turbulence statistics received a  huge attention since 1941, when \cite{Kolmogorov1941} argued that  small scales have the best prospect to exhibit universal properties.   This theory did not account for the internal intermittency, defined as strong fluctuations in space and time of the  local, instantaneous kinetic energy dissipation rate $\varepsilon$  \citep{Batchelor_Townsend_nature_1949, Townsend1951fine}.  
	While the famous Kolmogorov turbulence theory in 1941 accounted for neither the internal intermittency phenomenon nor the finite Reynolds number effect (FRN), e.g.  \cite{Shunlin2017, Shunlin2018},   Kolmogorov theory 1962  \citep{Kolmogorov1962} was underpinned by  modified similarity hypotheses, aimed at accounting for intermittency.   
	One important merit of K41 and K62 is that they confer a phenomenological \citep{Kolmogorov1941, Kolmogorov1962} and a theoretical \citep{Kolmogorov1941a} framework allowing to link statistics at large scales (presumably, within in an inertial range) and the smallest scales at which $\varepsilon$ is properly defined.    Numerous  later studies  \citep{She_1994universal,Shunlin2018, Yakhot2003, zhou_turbulence_2021, Shi_qian_2021}  discussed the inappropriateness of these hypotheses, and proposed adequate amendments. One of them is the accounting of the FRN effect, which implies to consider in theoretical developments all specific physical phenomenon of the flow, such as decay, diffusion, production, etc.  The approach developed in this work follows this philosophy, and considers all terms in the transport equations, none of them being a priori neglected. 
	
	Turbulence statistics which pertain to internal intermittency  usually encompass  two kinds of methods: i) one-point statistics of small scales (reflected by gradients of the velocity field); and ii)  two-point statistics, particularly by the scaling exponents of higher-order structure functions. Note that the small-scale limit of ii) fully recovers i).  Scaling laws of longitudinal structure functions of order $p$, defined as the difference of the velocity component $u$ between two space points separated by the scale $r$, are sought as:
	\begin{equation}
		\langle (u(x+r) - u(x))^p \rangle \sim r^{\zeta_p},
	\end{equation}
	where $u$ is the $x$-component velocity in the $(x,y,z)$  reference system, $r$  the separation distance between the two points and $\langle \rangle$ denotes averaging.  Assessing the scaling exponents demands particular care. 
	Strictly speaking, they can only be correctly assessed in a range of scales called  'inertial sub-range', which, in turn, requires a large Reynolds number. The exact value of the threshold depends on the flow: for instance,   \cite{Ishihara_2009review} showed that $Re_\lambda$ (based on Taylor's microscale $\lambda$) must exceed   $500$, which implies a minimum resolution of $1024$ in a periodic box simulating homogeneous and isotropic turbulence. This requirement is very impelling for the computational resources of DNS. 
	For lower Reynolds numbers, it is common to designate as Restricted Scaling Range (RSR) those scales for which a scaling of different statistics can be discerned. In the RSR, the value of the scaling  exponent is smaller than the asymptotic prediction of Kolmogorov.  K41  predicts that, under the assumption of sufficiently high Reynolds numbers,   the structure function of order $p$ should scale as $\zeta_p^K=p/3$ within the inertial range (the superscript  $K$ denotes 'Kolmogorov'). The prediction is exact for $p=3$ since the K\'arm\'an-Howarth-Kolmogorov equation is deduced from the Navier-Stokes equations and grants the 4/5 law for longitudinal 3rd-order structure function, for sufficiently high Reynolds numbers.  However,  for $p>3$, the deviation of the scaling exponent $\zeta_p$ from $p/3$ is often attributed to the  effect of internal intermittency, although the FRN effect is also mixed up with intermittency \citep{Shunlin2017}. For classical turbulence, a solid theory for predicting higher-order moments scaling laws is still missing. One of the intricacies stands in the correct account of the FRN, and associated closures for the numerous terms highlighted in  transport equations \citep{Shunlin2018, Shi_qian_2021, zhou_turbulence_2021}. 
	 Intermittency has also been addressed through GP models \citep{krstulovic2016}. It is outlined that 
the incompressible velocity are found to be skewed for turbulent states. Comparisons with  homogeneous and isotropic Taylor-Green flow, revealed  the universality of
the statistics, including a Kolmogorov constant
close to the one of classical fluid. 

	The  HVBK model of QT at finite temperature  it is the perfect framework to develop such statistical analysis, since the two components of the flow are governed by  Navier-Stokes type equations (over which the coupling, mutual friction term, is to be accounted for) and thus can be easily separated. The two components are denoted by subscripts 'n' and 's' standing for the normal fluid and superfluid, respectively. 	The total density of the fluid is  the sum of each component densities, $\rho=\rho_n+\rho_s$. The density ratio is temperature-dependent. 	For $T \approx T_\lambda$,  $\rho_n/\rho = 1$ and for $T = 0$, $\rho_n/\rho = 0$. Both experimental \citep{Rusaouen_etal_2017} and numerical (based on the HVBK shell model) \citep{Biferale_2018, Shukla2016, Lvov_2006} studies were devoted to inspecting intermittency by analyzing the scaling exponents for  higher-order structure functions.  A consensus emerged that the intermittency of quantum turbulence is very similar to classical turbulence for temperatures close to $T_\lambda$,  or  close to absolute zero (see Table I in \cite{Rusaouen_etal_2017}). There is no clear conclusion for  intermediate temperatures (between $T_\lambda$ and $0$). Experimental studies covered a wide range of temperatures ($0<T<T_\lambda$) and concluded that for quantum turbulence, the higher-order scaling exponents are smaller than $p/3$ as in classical turbulence, and they are almost unaffected by the temperature \citep{Rusaouen_etal_2017}. However, HVBK shell model studies lead to different conclusions at intermediate temperatures, where $\rho_n \approx \rho_s$.  \cite{Shukla2016} claim that for the quantum turbulence in intermediate temperatures $\rho_n \approx \rho_s$, the scaling exponents are more significant than the Kolmogorov prediction,  $\zeta_p^c<\zeta_p^K<\zeta_p^q$  (superscripts 'c' and 'q' stand for 'classical' and 'quantum',  respectively), while \cite{Boue_2013} found that scaling exponents are smaller than the Kolmogorov prediction and even smaller than the scaling exponents of classical turbulence,  $\zeta_p^q<\zeta_p^c<\zeta_p^K$. \cite{Biferale_2018}  performed DNS for a gradually damped HVBK model and provided support for the latter conclusion. This discrepancy is due to the additional effect of the mutual friction force  in both normal fluid and superfluid, in the case of $\rho_n \approx \rho_s$.

In the present work, we use DNS results based on the HVBK model for forced homogeneous isotropic turbulent flow with Reynolds number of the  normal fluid (based on Taylor's microscale) close to 100. We consider density ratios $\rho_n/\rho_s$ between 0.1 and 10, corresponding to temperature spanning $[0, T_\lambda]$.  Because of the moderate Reynolds numbers of the normal fluid, the range of scales over which statistics will be revealed are: the dissipative range, the RSR (intermediate scales), and large scales (comparable with the size of the simulation box, at which forcing is applied).   
The first question we address is  the role of the mutual friction in the transport equation of the  3rd-order structure function.   We deduce this  equation from the first principles (here, two-fluids HVBK) by accounting for the FRN  effect at each scale and different temperatures, as a function of the density ratio $\rho_n/\rho$. Each term of the balance equation is assessed from DNS data. We corroborate this analysis with 
one-point statistics of velocity derivatives, which is another tool to probe turbulent intermittency.  We quantitatively study the tails of Probability distribution functions (PDFs)  of velocity derivatives  by computing the flatness, defined as the 4th-order moment normalised by the square of the 2nd-order moment. We then compare with CT, for which DNS at very high Reynolds numbers \citep{Ishihara_2007} revealed that the flatness of the velocity gradients is much larger than $3$ (typical for a Gaussian distribution).
Despite the easy accessibility of small scales in  numerical simulations of QT,  we are not aware of any report of  similar analysis for probing internal intermittency.

The paper is organized as follows. Section \ref{section_HVBK} describes the two-fluids HVBK model and the main parameters of  direct numerical simulations.  Section \ref{section_SbS} is devoted to inspecting each term in the transport equation of the 3rd-order structure-function,  with particular attention paid to the influence of the mutual friction term over the whole range of scales and for different density ratios. Section \ref{section_onepoint} reports one-point statistics of the longitudinal velocity gradients of each fluid component and the total velocity of the turbulent flow. Section \ref{section_flatnessderivative} deals with the flatness of the velocity derivative. Conclusions are drawn in Sect. \ref{section_conclusions}.


\section{The HVBK model and Direct Numerical Simulations} \label{section_HVBK}

We use the so-called incompressible  Hall--Vinen--Bekharevich--Khalatnikov (HVBK)  model \citep{QT-Lipniacki-2006,2009_Donnelly_yearophysic}.  Navier-Stokes equations describe the normal fluid (variables with subscript 'n') and the superfluid motion (variables with subscript 's') is governed by an Euler-like equation:
\begin{eqnarray}
\nabla\cdot  \bm{v}_n =0 , \; \nabla\cdot \bm{v}_s =0,
\label{eq:HVBK-continuty}\\
\frac{\partial\bm{v}_n}{\partial t}+(\bm{v}_n \cdot \nabla)\bm{v}_n=-\frac{1}{\rho_n}\nabla p_n  + \frac{1}{\rho_n} \bm{F}_{ns} + \nu_n \nabla^2\bm{v}_n,
\label{eq:HVBK-n}\\
\vspace{0.2cm}
\frac{\partial\bm{v}_s}{\partial t}+(\bm{v}_s \cdot \nabla)\bm{v}_s=-\frac{1}{\rho_s}\nabla p_s  - \frac{1}{\rho_s}\bm{F}_{ns}+ \nu_s \nabla^2\bm{v}_s, 
\label{eq:HVBK-s} 
\end{eqnarray}
where $\nabla$ stands for the nabla operator, $\bm{v}$ is  the velocity vector, $\rho $ the density and $p$ the pressure.   Note that the superfluid viscosity $\nu_s$ is theoretically zero, and it is added for the purpose of stability of numerical simulations at very low temperatures.  It may also be viewed as a crude surrogate for the superfluid dissipation processes at inter-vortex scales and below.

The two fluid components are coupled through a mutual friction force $\bm{F}_{ns}$. The form of the friction force is  \citep{Hall_Vinen_1956},  \cite{Lvov_2006}:
\begin{equation}
\bm{F}_{ns} =  -\frac{B}{2}\frac{\rho_s\rho_n}{\rho}\frac{\bm{\omega}_s \prodvec (\bm{\omega}_s  \prodvec (\bm{v}_s -\bm{v}_n))}{|\bm{\omega}_s |}   -\frac{B'}{2}\frac{\rho_s\rho_n} {\rho}\bm{\omega}_s  \prodvec (\bm{v}_s -\bm{v}_n),
\label{eq:FM-simple0}
\end{equation}
where $\bm{\omega}_s = \nabla \prodvec\bm{v}_s$ is the coarse-grained superfluid vorticity (see below). 
We assume that for the superfluid the predominant energy loss is due to macroscopic friction with the normal fluid. We implicitly neglect dissipation process by vortex reconnection. This certainly excludes the validity of such a model for temperatures very close to $0K$, and does not allow to investigate scales smaller than the intervortex distance.    The perpendicular component of the force in Eq.  \eqref{eq:FM-simple0} is neglected, since it does not contribute to the energy exchange.  A discussion on the impact of these simplifications is provided in Appendix \ref{appendix_mutualfriction}.
 The simplified form of the friction force is then \citep{Lvov_2006}:
\begin{equation}
\bm{F}_{ns} = -\frac{B}{2}\frac{\rho_s\rho_n}{\rho}|\nabla \prodvec\bm{v}_s|  (\bm{v}_n -\bm{v}_s),
\label{eq:FM-simple}
\end{equation}
where $B$ is a temperature related parameter, measured in various experiments (see for instance  \cite{Barenghi_1983}). We set the value $B=1.5$ corresponding to the averaged value extracted from experimental data. 

  This calculation of the mutual friction was based on Feynman's rule. Assuming that a large number of superfluid  vortices of quantized circulation $\kappa$ are parallel (polarized) in a bucket, the equivalent  solid-body rotation vorticity is $2 \bm{\Omega}=|\nabla \prodvec\bm{v}_s| = |\bm{\omega}_s |=\kappa \mathcal{L}$, where $\mathcal{L}$ is the vortex line density per unit volume and $\bm{\Omega}$ the equivalent angular velocity.  The equivalent averaged  coarse-grained velocity of the superfluid is then $\bm{v}_s = \bm{\Omega} \times \bm{r}$. 
  The validity of  the expression of the mutual friction force \eqref{eq:FM-simple}  in general quantum turbulent flows, where vortex lines are randomly oriented rather than highly polarized, is still matter of debate. The existence in QT of dense vortex clusters (bundles) with quasi-parallel vortex lines \citep{Narimsa_2011,Baggaley_2012,galantucci2021-bundle} supports the idea of an averaged vorticity. Obtaining a model equation for the evolution of  $\cal L$ that accounts for non-polarized vortices is still an open question \citep{QT-Lipniacki-2006,nemirovskii2020}.

The kinematic viscosity $\nu_n =\mu/\rho_n$ in Eq. \eqref{eq:HVBK-n} is a simulation parameter. Based on the concept of the two-fluid model, $\rho_n$ decreases with temperature, while  the dynamic viscosity $\mu$ is also temperature-dependent. Naturally, the parameter $\mu$ in the two-fluid model should be taken as the dynamic viscosity $\mu_*$, which was measured in superfluid helium for a range of temperatures $1K<T<T_\lambda$ (see \cite{Barenghi_1983}). It is common practice in HVBK simulations to fix $\mu$ as a constant, independent of temperature. We adopt this simplification, since the dynamic viscosity of the normal fluid could be different from $\mu_*$ at low temperatures because of other dissipative effects in the superfluid. We choose here to fix $\nu_n$  as a constant, independent of the temperature.

We solved numerically the system of equations \eqref{eq:HVBK-continuty}-\eqref{eq:HVBK-s} using Fourier pseudo-spectral methods classically used for Navier-Stokes equations. Direct numerical simulations were performed by adapting a Navier-Stokes code that proved  efficient and accurate in computing high-order statistics of turbulent flows \citep{gauding2017high}.  Periodic boundary conditions were applied to a computational box of  length $2\pi$. Grid resolution was $512^3$, which was sufficient to reach a moderate $Re_\lambda\sim 100$, based on Taylor's microscale.  We have also performed numerical simulations with a better resolution of $1024^3$ (see Appendix \ref{appendix_resolution}).  The results reported are not affected by the resolution, except the value of the flatness of the velocity derivative of the superfluid, as discussed later.  To achieve a quasi-stationary homogeneous isotropic turbulence, an additional forcing term was added in the momentum equations \eqref{eq:HVBK-n} and \eqref{eq:HVBK-s} at large scales.

The energy injection rate $\varepsilon_{*}$ 
 is constant in time,  for different temperatures, and for  both fluid components.   We set  $\varepsilon_*=7e-4$  for all simulations. The energy injected in superfluid is transferred by mutual friction and eventually dissipated by the normal fluid component. But, the energy transfer becomes less efficient for low temperatures because of $\rho_n/\rho$ tending to zero. 
 Accounting for an additional forcing term would result in unstable simulations. To maintain the stability of the simulations for very low temperatures, a common technique in the HVBK model is to impose an artificial viscosity $\nu_s$ to the superfluid.  To respect the two-fluid concept, one should make sure that the artificial viscosity of the superfluid is much smaller than the viscosity of the normal fluid,  $\nu_s\ll\nu_n$.

\begin{table}
	\centering
	\begin{tabular}{c c c c c c c c}
		\hline
		$\rho_n/\rho_s$ & $Re_H$ & $Re_\lambda$ & $\tau_L/\tau_\eta$& $\frac{\overline{\varepsilon} H}{\mathcal{K}^{3/2}}$ & $L/\eta$ & $\eta/\Delta$ &$\delta/\Delta$\\
		\hline
		0.91 & 2.21e+3 & 90.75  & 12.54  & 2.17 & 60.70 & 1.120 & -\\
		0.74 & 2.23e+3 & 85.12  & 12.91  & 2.51 & 60.54 & 1.078 & -\\
		0.55 & 2.22e+3 & 80.47  & 13.25  & 2.82 & 60.41 & 1.0428 & -\\
		0.50 & 2.20e+3 & 74.11  & 14.99  & 3.27 & 65.55 & 1.019 & -\\
		0.43 & 2.18e+3 & 71.97  & 15.35  & 3.44 & 66.18 & 1.011 & -\\
		0.19 & 2.19e+3 & 58.01  & 18.33  & 5.32 & 70.95 & 0.903 & -\\
		0.09 & 2.23e+3 & 55.17  & 20.69  & 5.98 & 78.08 & 0.867 & -\\
		\hline
		0.91 & 2.24e+4 & 786.54 & 14.348 & 0.296 & 204.47 & 0.326 & 2.30\\
		0.74 & 2.26e+4 & 722.87 & 15.068 & 0.354 & 205.86 & 0.309 & 2.19\\
		0.55 & 2.26e+4 & 737.24 & 14.477 & 0.339 & 199.73 & 0.313 & 2.06\\
		0.50 & 2.23e+4 & 600.66 & 18.277 & 0.506 & 227.61 & 0.285 & 2.01\\
		0.43 & 2.22e+4 & 577.96 & 18.890 & 0.543 & 230.76 & 0.281 & 1.95\\
		0.19 & 2.24e+4 & 421.78 & 24.838 & 1.027 & 259.20 & 0.239 & 1.70\\
		0.09 & 2.27e+4 & 376.28 & 29.858 & 1.310 & 294.31 & 0.220 & 1.57\\
		\hline
	\end{tabular}
	\caption{Simulation parameters of the turbulent flow field. $H = 2\pi$ is the size of the numerical domain and $\Delta =H/512$  the mesh size in each direction. $Re_H$ and  $Re_\lambda$ are the Reynolds numbers based on the large scale of the flow and Taylor's microscale, respectively. $\mathcal{K}$ is the mean turbulent kinetic energy and $\varepsilon$ the mean energy  dissipation rate. 
The eddy turnover time was computed as 
		 $\tau_L = (2/3\mathcal{K})/\varepsilon$  and the  scale of the large eddies as $L = (2/3 \mathcal{K})^{3/2} /\varepsilon$.  
		  Kolmogorov length and time scales are $\eta$ and $\tau_\eta$, respectively.  $\delta = {\cal L}^{-1/2}=\sqrt{\kappa/|\omega_s|}$ is the inter-vortex length scale, with $\kappa=\nu_n(\kappa_{phys}/\nu_{phys})$. In all computations $\kappa_{phys}\approx 1e-7(m^2/s)$ and $\nu_{phys}\approx 2.0e-8(m^2/s)$. }
	\label{table_param-II}
\end{table}

To summarize, in the present work we fix $\nu_n$ as a constant independent of the  temperature, and $\nu_s=0.1\nu_n$ for all numerical simulations. Other techniques exist, like the gradually damped HVBK model \citep{Biferale_2018} and the shell model  \citep{Boue_2015}, for which a temperature-dependent normal fluid viscosity $\nu_n$ and temperature-dependent superfluid viscosity $\nu_s$ are imposed. The statistics of the DNS HVBK model were computed over $30$ integration time scales. Table \ref{table_param-II} contains simulation parameters for all 7 considered cases.  The first part refers to the normal fluid, while the second one pertains to the superfluid.


\section{Scale-by-scale evolution of the third and fourth-order structure functions of the normal and superfluid} \label{section_SbS}


We present in this section the scale-by-scale budget equation for the 4th-order structure functions  of velocity increments in a HVBK turbulent flow. We start from the transport equation of the 3rd-order structure function for a single-fluid Navier-Stokes turbulent flow. This  equation was derived by  \cite{Hill2001}  and \cite{Yakhot2003} and assessed through experimental and numerical data by \cite{Hill_Boratav2001}, and \cite{Gotoh_Nakano_2003}.  Denoting by  $r$ the space increment, $\delta u=u(x+r)-u(x)$ the longitudinal velocity increment and $\delta v=v(x+r)-v(x)$ the transverse velocity increment, the following transport equation was established under the assumption of homogeneity and isotropy:
\begin{equation}
\underbrace{\partial_t D_{111}}_{Term1} + \underbrace{\left(\partial_r+ \frac{2}{r}\right) D_{1111}}_{Term2} - \underbrace{\frac{6}{r} D_{1122}}_{Term2'}= \underbrace{-T_{111}}_{Term3}+ \underbrace{2\nu C}_{Term4}- \underbrace{2 \nu Z_{111}}_{Term5},  \label{eq1}
\end{equation}
where $\partial_r \equiv \partial/ \partial r$, $\nu$ is the kinematic viscosity and
\begin{align}
D_{111}&=\langle{(\delta u)^3}\rangle, \nonumber \\
D_{1111}&=\langle{(\delta u)^4}\rangle, \nonumber \\
D_{1122}&=\langle{(\delta u)^2(\delta v)^2}\rangle, \nonumber \\  
C(r,t)&=-\frac{4}{r^2} D_{111} (r,t) + \frac{4}{r} \partial_r D_{111} + \partial_r \partial_r D_{111}, \nonumber \\
Z_{111}&=3 \left\langle{\delta u \left[\left(\frac{\partial u}{\partial x_l} \right)^2+\left(\frac{\partial u'}{\partial x_l'} \right)^2\right]}\right\rangle. \label{eq2}
\end{align}
In expressions \eqref{eq2} double indices indicate summation (over $l=1,2,3$) and a prime refers to variables at point $x+r$. Term $Z_{111}$, also  called {\em dissipation-source term}, couples components of the dissipation with $\delta u$, and thus acts at all scales (this will be discussed in detail later). Term $T_{111}$ is related to the pressure gradient and has the form:
\begin{equation}
{T_{111}=3\left\langle{\left( \delta u\right)^2 \delta \left( \frac{\partial p}{\partial x}\right)}\right\rangle}. \label{eq3}
\end{equation}

We apply the same approach to obtain a similar transport equation for  HVBK equations \eqref{eq:HVBK-n}-\eqref{eq:HVBK-s}, which have Navier-Stokes  structure.
We denote by $D_{111}^n$ and  $D_{111}^s$ the  3rd-order longitudinal structure functions for normal and superfluid components, respectively.  The two transport equations are:
\begin{align}\nonumber
\underbrace{\partial_t D_{111}^n}_{Term1} + \underbrace{\left(\partial_r+ \frac{2}{r}\right) D_{1111}^n}_{Term2} + \underbrace{\left(-\frac{6}{r} D_{1122}^n \right)}_{Term2'}=& \underbrace{-T_{111}^n}_{Term3}+ \underbrace{2\nu_n C^n}_{Term4} +\underbrace{\left(-2 \nu_n Z_{111}^n\right)}_{Term5}\\ &+ \underbrace{\langle{(\delta u_n)^2 (3 \frac{\rho_s}{\rho}\delta F_{\parallel}^{ns})}\rangle}_{Term 6}+\underbrace{3 \langle{(\delta u_n)^2 \delta f_{\parallel}^{n}}\rangle}_{Term 7} ,  \label{eq:D3_balance_n}
\end{align}
\begin{align}\nonumber
\underbrace{\partial_t D_{111}^s}_{Term1} + \underbrace{\left(\partial_r+ \frac{2}{r}\right) D_{1111}^s}_{Term2} + \underbrace{\left(-\frac{6}{r} D_{1122}^s\right)}_{Term2'}=& \underbrace{-T_{111}^s}_{Term3}+ \underbrace{2\nu_s C^n}_{Term4} + \underbrace{\left(-2 \nu_s Z_{111}^s\right)}_{Term5}\\&+ \underbrace{\langle{(\delta u_s)^2 (-3 \frac{\rho_n}{\rho}\delta F_{\parallel}^{ns})}\rangle}_{Term 6}+\underbrace{3 \langle{(\delta u_s)^2 \delta f_{\parallel}^{s}}\rangle}_{Term 7}.  \label{eq:D3_balance_s}
\end{align}
For the sake of simplicity, we used the same notations for different terms as in Eq. \eqref{eq1}, while referring to either normal or  superfluid components.  New \textbf{\textit{Term6}} and \textbf{\textit{Term7}} appear. The former  comes from the mutual friction force $\bm{F}_{ns}$ (appearing with opposite signs in the two equations) and the latter from  forcing terms $\bm{f}^{n}$ and $\bm{f}^{s}$ added in both equations to force turbulence.

Equations \eqref{eq:D3_balance_n} and  \eqref{eq:D3_balance_s} allow us to obtain exact expressions of the 4th-order structure function (and, further on, of the flatness factor). Recalling that  $\left(\partial_r+ {2}/{r}\right)=\left(\partial_r(r^2 )\right)/r^2$, we obtain after integration with respect to the scale $r$: 
\begin{equation}
D_{1111}^n=\frac{1}{r^2} \int_0^r s^2 \left(-Term1-Term2'+Term3+Term4+Term5+Term6+Term7\right)^n  ds,  \label{eq:D4_n}
\end{equation}
\begin{equation}
D_{1111}^s=\frac{1}{r^2} \int_0^r s^2 \left(-Term1-Term2'+Term3+Term4+Term5+Term6+Term7\right)^s  ds.  \label{eq:D4_s}
\end{equation}

\begin{figure}
			\begin{center}
	\begin{minipage}{0.65\textwidth}
			\includegraphics[width=\textwidth,keepaspectratio,trim={1.2cm 7.2cm 1.4cm 7.5cm}, clip]{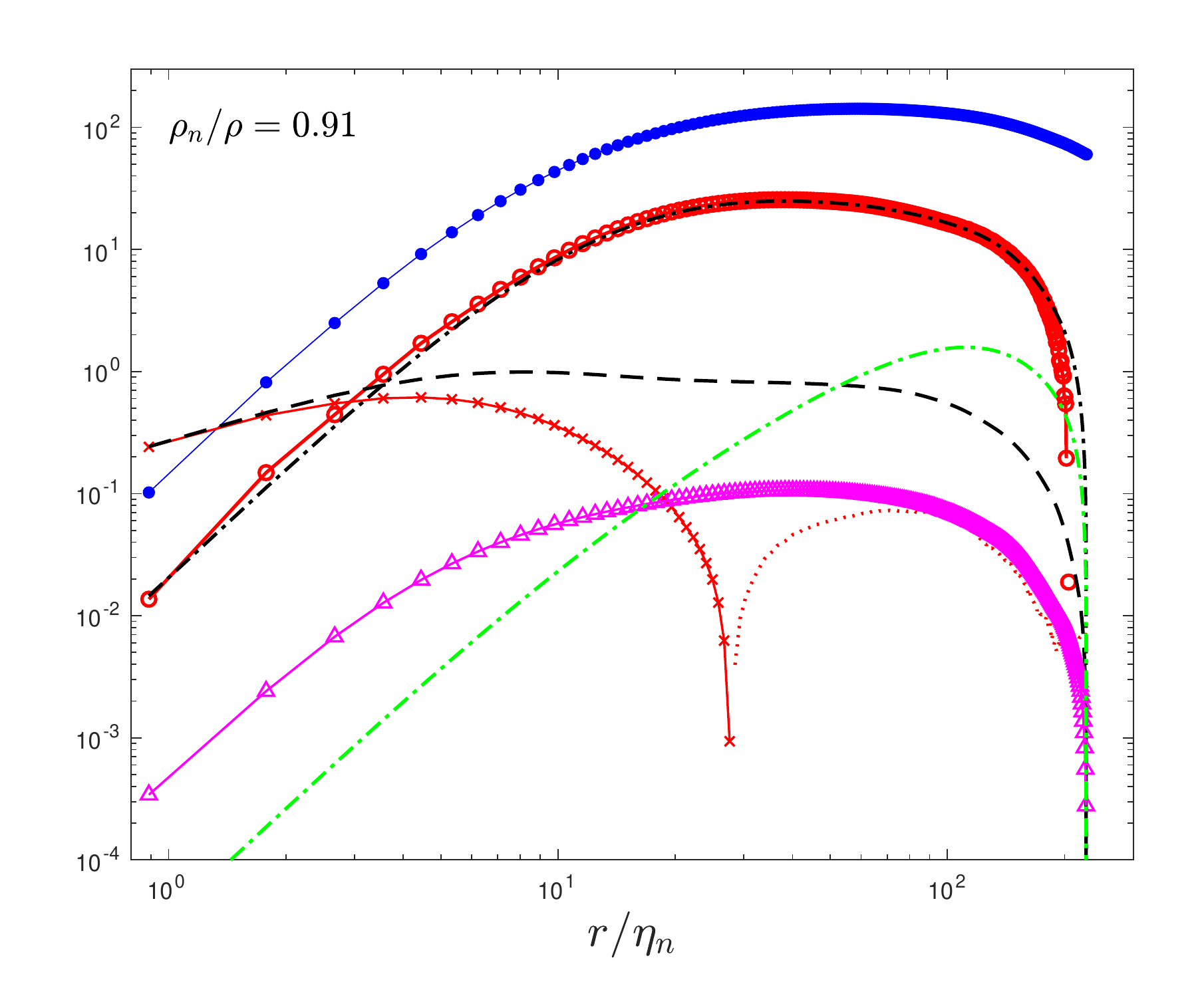}
	\end{minipage}
	\begin{minipage}{0.65\textwidth}
					\begin{tabular}{|l|l|l|l|}
				\hline
				Symbol & Colour & Terms &\\
				\hline
				$\bullet -$ & blue & $(\partial_r +2/r)D_{1111}$ & Term2\\
				\hline
				$\circ -$  & red & $(\partial_r +2/r)D_{1111}-\frac{6}{r} D_{1122}$& Term1+Term2\\
				\hline
				$- \cdot$ & black & $-T_{111}$ & Term3\\
				\hline
				$\times -$& red & 	$-2\nu C$ & -Term4\\
				\hline
				$\cdot \cdot \cdot$& red & 	positive part of $2\nu C$ & (Term4)$_+$\\
				\hline
				$- \  -$  & black & $-2\nu Z_{111}$ &Term5 \\
				\hline
				$\triangle -$ & magenta & $-3 \frac{\rho_s}{\rho}\langle{(\delta u_n)^2 \delta F_{\parallel}^{ns}}\rangle$ & -Term6 \\
				\hline
				$- \cdot$&green  & $3 \langle{(\delta u_n)^2 \delta f_{\parallel}^{n}}\rangle$ & Term7\\
				\hline
			\end{tabular}
	\end{minipage}
		\end{center}

	\caption{Case $\rho_n/\rho=0.91$ (the normal fluid is dominant). Terms in the budget Eq. \eqref{eq:D3_balance_n} for the normal fluid.   Scale $r$ is  normalized by  the Kolmogorov scale $\eta_n$ of the normal fluid. All terms are normalized by $\varepsilon_*^{5/4}\nu_n^{1/4}$, with $\varepsilon_*=7$e-4 the constant energy rate injected to force turbulence.}
	\label{fig:D3-balance-normal-rho10-complete}
\end{figure}

To assess the importance of each term in transport equations \eqref{eq:D3_balance_n}-\eqref{eq:D3_balance_s} for the 3rd-order structure functions, we naturally start with the simulation case $\rho_n/\rho=0.91$ (see Tab. \ref{table_param-II}). For this case, where the normal fluid is predominant,  the results are expected to be similar to those known for a classical single-fluid turbulent flow \citep{Hill_Boratav2001}.  Figure  \ref{fig:D3-balance-normal-rho10-complete} shows the scale-dependence of each term in  Eq. \eqref{eq:D3_balance_n}, after normalization by $\varepsilon_*^{5/4}\nu_n^{1/4}$, with $\varepsilon_*$ the  constant energy rate injected to force turbulence.  
Note that for this case the smallest resolved scale is smaller than the Kolmogorov scale $\eta_n$ (see Tab. \ref{table_param-II}). 

  \textbf{\textit{Term1}}  reflects the temporal decay of the 3rd-order structure function. This term is zero for steady-state flows and thus negligible here. \textbf{\textit{Term2}}  (blue $\bullet -$) is the prevalent term over the scales within the RSR. The sum of the two transport terms \textbf{\textit{Term2}}+\textbf{\textit{Term2'}}  (red $\circ$) balances the pressure-related  \textbf{\textit{Term3}} (black $- \cdot$ ) fairly well, over the whole range of scales. 
 
  \textbf{\textit{Term4}} (red $\times-$, plotted with changed sign) is negative at small scales and positive at large scales. It represents the viscous destruction of the 3rd--order structure functions. As expected, this contribution is negligible over the RSR, but becomes important in the viscous range. At the smallest resolved scale, this is the most prevalent term and  is balanced by 
 \textbf{\textit{Term5}} (black ---), the dissipation source term. This term exhibits a plateau over the RSR, and it is $15 \%$ of the other terms. Albeit smaller, this term cannot be ruled out. 
 
   \textbf{\textit{Term6}} (magenta $\triangle -$), representing the friction force coupling,  is the less important term. This seems reasonable behaviour for a fluid essentially composed of normal fluid.   Note also that \textbf{\textit{Term6}} is negative for the normal fluid, so  the figure illustrates (-\textbf{\textit{Term6}}).  Finally,  the forcing term (\textbf{\textit{Term7}}) (green $- \cdot$)  affects the very large scales only and its effect gradually diminishes towards small scales. 

The analysis of this case suggests that, as in classical single-fluid turbulence \citep{Hill_Boratav2001},  the two transport terms (\textbf{\textit{Term2}}+ \textbf{\textit{Term2'}}) are only balanced by the pressure-related term \textbf{\textit{Term3}}. This occurs over the whole range of scales, albeit the effect of the viscosity is obviously felt within the viscous range. The same conclusion was reached by
 \cite{Hill_Boratav2001} on the basis of experimental and DNS data. However, these authors did not calculate exactly the dissipation source term, neither the forcing term (which was neglected within the derivation, on the basis of the assumption of  very large Reynolds numbers). They also noted departures from homogeneity and isotropy, which are clearly observed in our simulations. 

Another important remark is that, despite the low Reynolds number of the flow ($R_\lambda \leq 100$), all terms that might have represented the FRN effect (friction force coupling through \textbf{\textit{Term6}}, forcing term \textbf{\textit{Term7}} and dissipation source term \textbf{\textit{Term5}}) are negligible. Therefore, there is no direct imprint of the FRN effect on the 4th-order moments of velocity increments. There is the possibility that this effect might be indirect, through the pressure field.   The conclusion that FRN effect is negligible is further comforted by other simulations for different temperatures (see below). The consequences are that 4th-order structure functions are only shaped by the pressure field. This observation was revealed by e.g. \cite{Yakhot2003,Gotoh_Nakano_2003}. The latter authors suggested a valuable model for the role of the pressure in turbulence.

\begin{figure}
	\begin{subfigure}{0.5\textwidth}
		\includegraphics[width=1.\textwidth,height=0.7\textheight,keepaspectratio,trim={1.5cm 7.2cm 2.4cm 8.5cm}, clip]{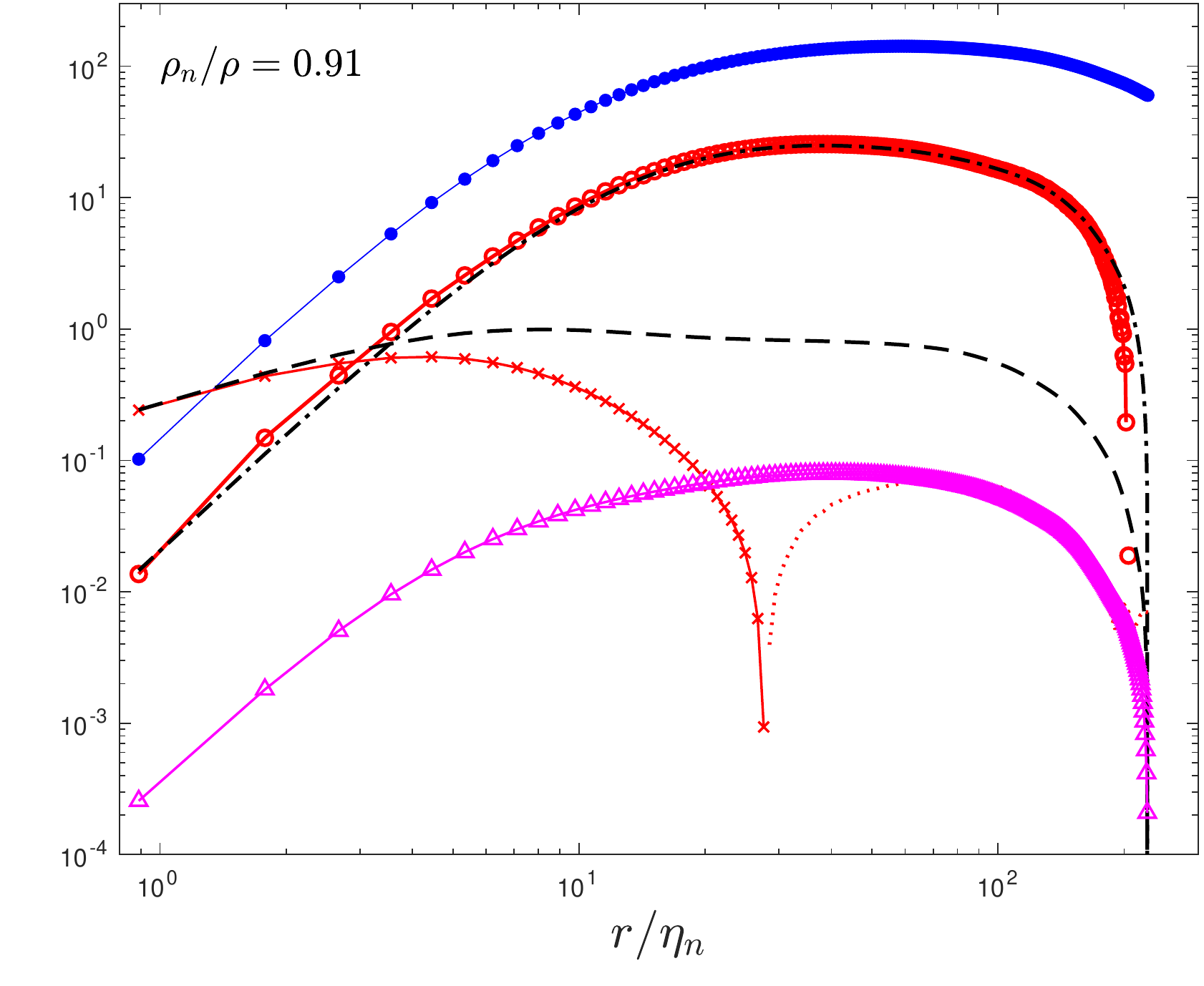}
		\caption{ }
		\label{fig:D3-balance-normal-rho10}
	\end{subfigure}%
	\hfill
	\begin{subfigure}{0.5\textwidth}
		\includegraphics[width=1.\textwidth,height=0.7\textheight,keepaspectratio,trim={1.5cm 7.2cm 2.4cm 8.5cm}, clip]{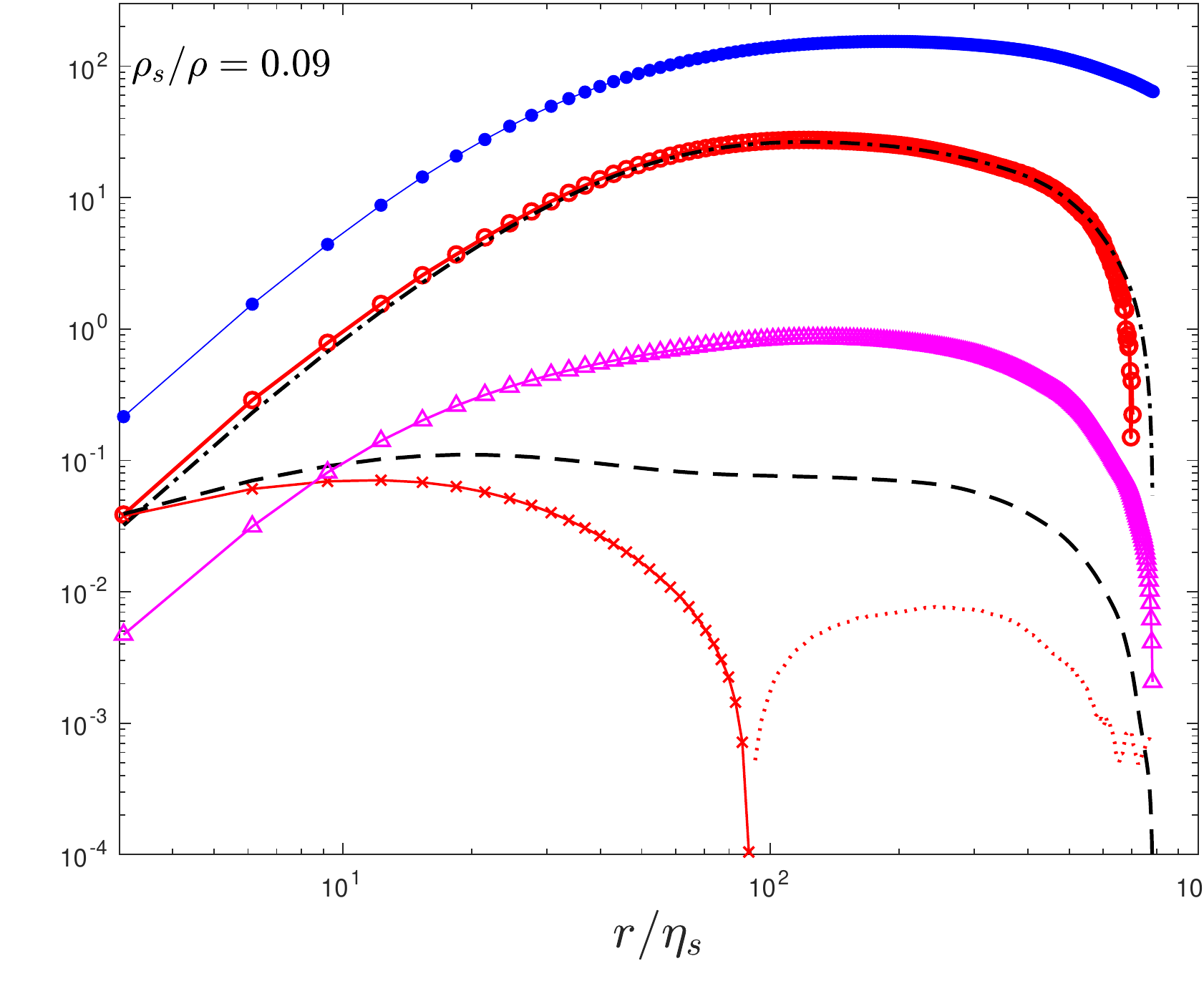}
		\caption{ }
		\label{fig:D3-balance-super-rho10}
	\end{subfigure}%

	\begin{subfigure}{0.5\textwidth}
		\includegraphics[width=1.\textwidth,height=0.7\textheight,keepaspectratio,trim={1.5cm 7.2cm 2.4cm 8.5cm}, clip]{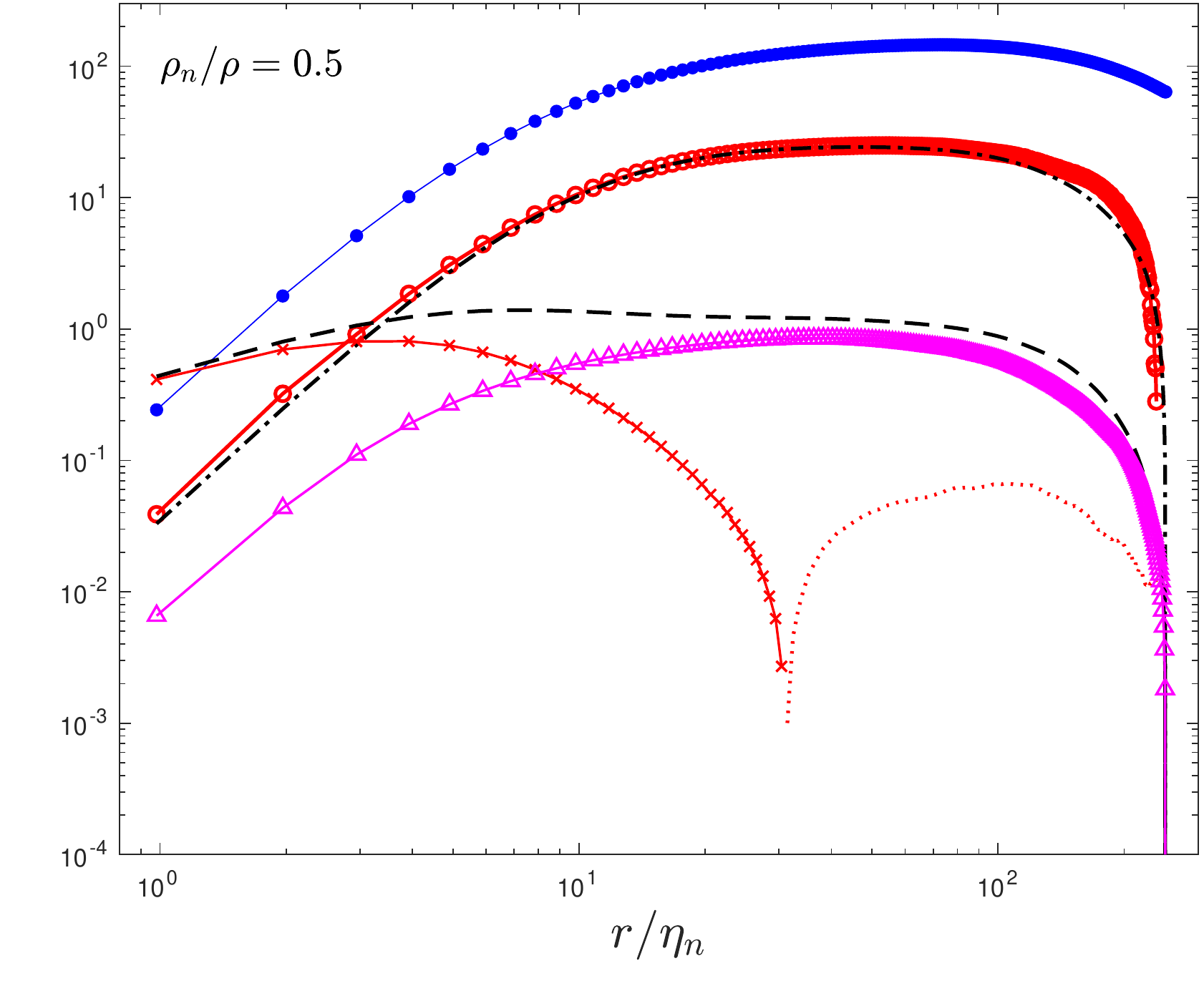}
		\caption{ }
		\label{fig:D3-balance-normal-rho1}
	\end{subfigure}%
	\hfill
	\begin{subfigure}{0.5\textwidth}
		\includegraphics[width=1.\textwidth,height=0.7\textheight,keepaspectratio,trim={1.5cm 7.2cm 2.4cm 8.5cm}, clip]{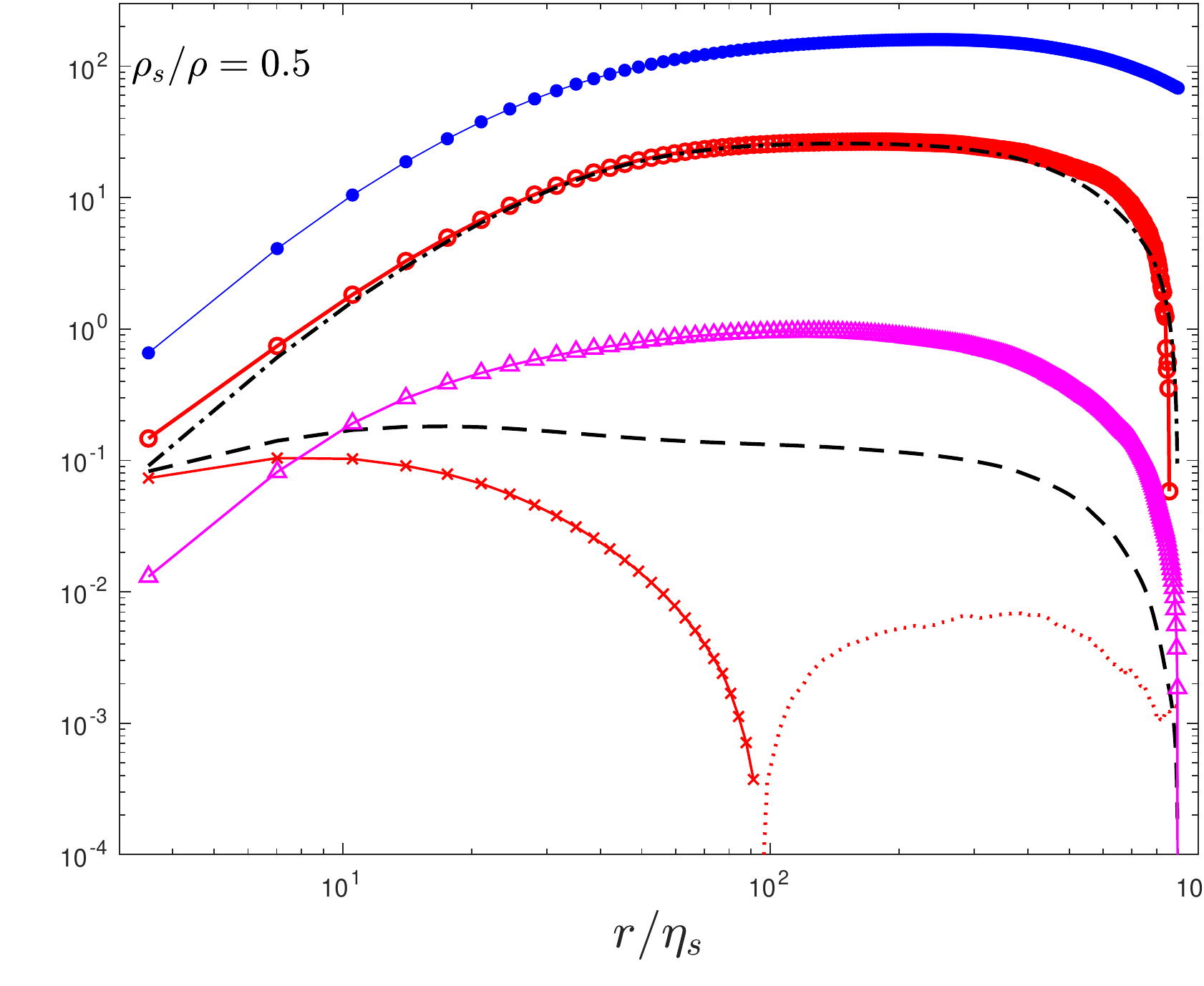}
		\caption{ }
		\label{fig:D3-balance-super-rho1}
	\end{subfigure}%

	\begin{subfigure}{0.5\textwidth}
		\includegraphics[width=1.\textwidth,height=0.7\textheight,keepaspectratio,trim={1.5cm 7.2cm 2.4cm 8.5cm}, clip]{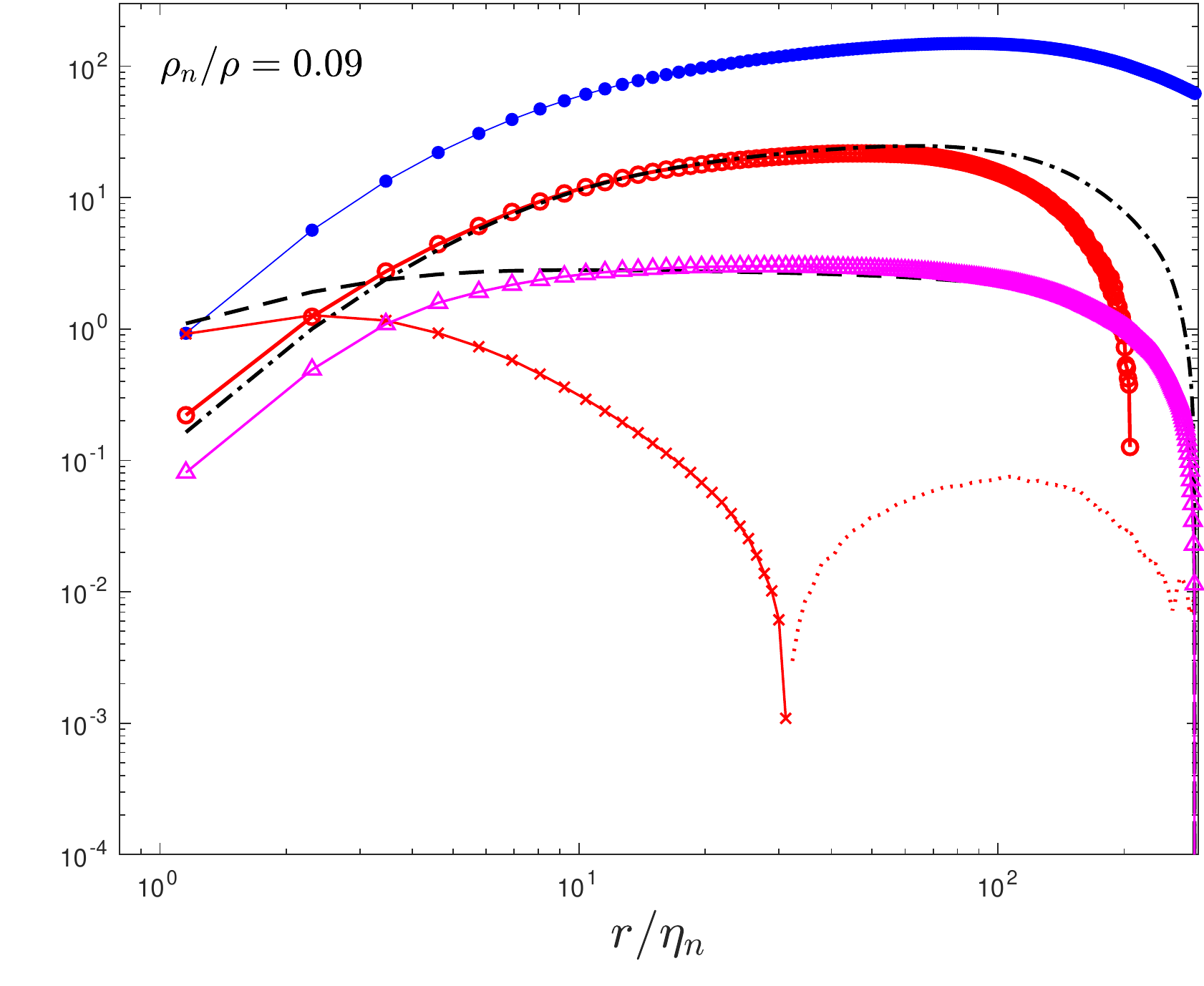}
		\caption{ }
		\label{fig:D3-balance-normal-rho01}
	\end{subfigure}%
	\hfill
	\begin{subfigure}{0.5\textwidth}
		\includegraphics[width=1.\textwidth,height=0.7\textheight,keepaspectratio,trim={1.5cm 7.2cm 2.4cm 8.5cm}, clip]{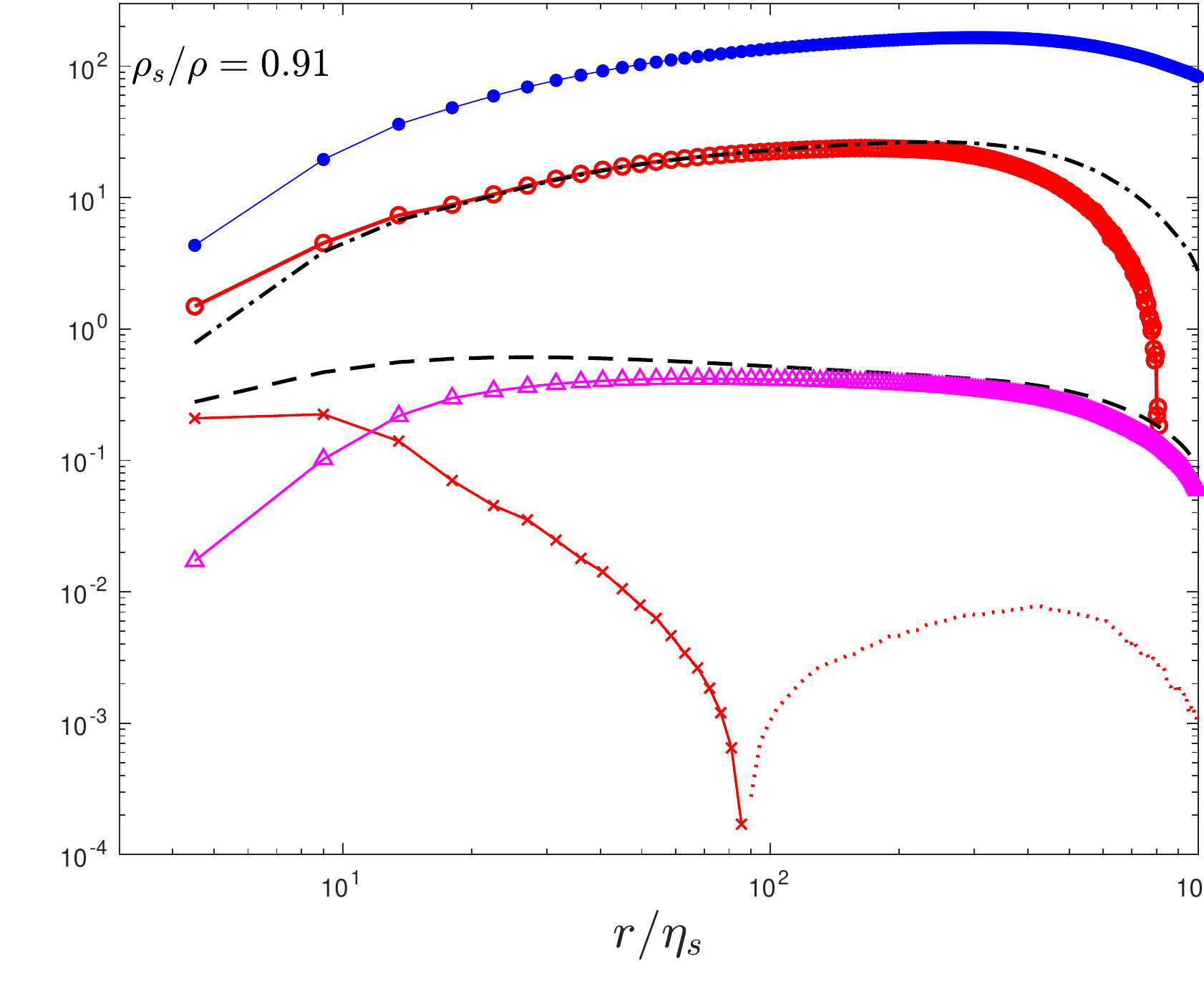}
		\caption{ }
		\label{fig:D3-balance-super-rho01}
	\end{subfigure}%
	\caption{Terms in budget Eq. \eqref{eq:D3_balance_n} for the normal fluid  (left column) and in Eq. \eqref{eq:D3_balance_s} for the superfluid  (right column).   Simulations for three density ratios $\rho_n/\rho = 0.91$ (a, b), $0.5$ (c, d) and $0.09$ (e, f). 
		Scale $r$ is normalized by Kolmogorov scales $\eta_n$ (normal fluid) and $\eta_s$ (superfluid), respectively. All terms are normalized by $\varepsilon_*^{5/4}\nu_n^{1/4}$, with $\varepsilon_*=7$e-4 the constant energy rate injected to force turbulence in both fluid fractions.  Same legend as in Fig. \ref{fig:D3-balance-normal-rho10-complete} for the graphical representation of different terms.}
	\label{fig:D3-balance}
\end{figure}

We now extend our analysis to other cases (see Tab. \ref{table_param-II}). We consider the case $\rho_n/\rho=0.5$ (temperature around $2K$) with balanced normal and superfluid fractions and the case $\rho_n/\rho=0.09$ (temperature close to 0.3K), with the superfluid dominating the flow. Terms in Eqs. 
\eqref{eq:D3_balance_n}-\eqref{eq:D3_balance_s} are depicted in Fig. \ref{fig:D3-balance}. For the coherence of the message, we replot in  upper panels of Fig. \ref{fig:D3-balance} the results obtained for $\rho_n/\rho=0.91$.

\textbf{\textit{Term1}}  reflects the temporal decay of the 3rd-order structure function. As stated above,  this term is absent in our simulations. We have kept it in the transport equations, as it provides a way to assess the degree at which other terms influence its behaviour. For a direct cascade, $D_{111}$ is negative. An enhance of the cascade is consistent with positive values of the  temporal derivative of  $(-D_{111})$. For the normal fluid, this enhancement can be the result of the friction force coupling, via  \textbf{\textit{Term6}}, which is negative (so -\textbf{\textit{Term6}} is positive).  Therefore, the cascade of the normal fluid may be enhanced by \textbf{\textit{Term6}}. The opposite effect stands for the superfluid, for which \textbf{\textit{Term6}} is positive. The origin of this different sign is at the level of the HVBK model, for which the coupling term is accounted for with different signs, reflecting an enhancement of the momentum for the normal fluid, and a reduction of the momentum for the superfluid.  
 
Forcing terms (\textbf{\textit{Term7}})  are not shown in Fig. \ref{fig:D3-balance}, because they only affect very large scales.
Generally speaking, as already emphasized, they exhibit similar behaviour to  classical turbulence, if  high temperatures are considered, corresponding to $\rho_n/\rho=1$. However, the additional mutual friction term \textbf{\textit{Term6}} plays a requisite role particularly for low temperatures, thus distinguishing the HVBK flow from classical fluids.
In the following section, we analyse  the results for each specific range of scales.

\clearpage
\clearpage

\subsection{Dissipative scales}
For the normal fluid, similar to classical turbulence at small scales, the pressure source  \textbf{\textit{Term3}} and   transport terms (\textbf{\textit{Term2}} + \textbf{\textit{Term2'}}) scale as  $r^3$.  In contrast, the viscous \textbf{\textit{Term4}} and dissipation source  \textbf{\textit{Term5}} vary proportionally to $r$. The viscous \textbf{\textit{Term4}} balances the dissipation source  \textbf{\textit{Term5}} for the very small scales. Although small differences  between these two terms are noticeable  for the lowest density ratio  ($\rho_n/\rho=0.09$, see Figs. \ref{fig:D3-balance-normal-rho01} and \ref{fig:D3-balance-super-rho01}), they are most likely due to the limited grid resolution. If we decrease $r$ to very small values, the two terms  eventually cancel each other. 
Moreover, transport terms (\textbf{\textit{Term2}} + \textbf{\textit{Term2'}}) are nearly balanced by the pressure source  \textbf{\textit{Term3}} for the normal fluid (as already discussed and illustrated in the left column of Fig. \ref{fig:D3-balance}).

For the superfluid (right column of Fig. \ref{fig:D3-balance}), unlike classical turbulence at smallest scales,  transport terms are slightly larger than the pressure source term. This difference is most likely attributable to  the equipartition of energy \citep{Salort_etal_2010}, which  finally results in the accumulation of energy at highest wavenumbers due to the very small value of the supefluid viscosity. 
Moreover, when the superfluid is dominant in the flow, the kinetic energy cannot be completely dissipated. This energy accumulates at the scales of the same order as the inter-vortex scale, which leads to an upward trend for the superfluid velocity spectrum. In quantum physics, this is associated with a partial thermalisation of superfluid excitations \citep{Barenghi_etal_2014}. Note that the upward trend of the superfluid velocity spectrum depends on  simulation parameters of the HVBK model. The truncated HVBK model resolves two coupled viscous fluids with different, albeit constant,  viscosities. The ability to settle the smallest scales of both fluids requires, nonetheless, a sufficiently high resolution. 

For small scales, the mutual friction term \textbf{\textit{Term6}} scales as  $r^3$ in both normal and superfluid components. \textbf{\textit{Term6}} decreases much faster than both  dissipation source and viscous terms. This underlines that  at small scales, the viscous and the dissipation source terms (both directly depending on the viscosity) are dominant.

\subsection{Intermediate scales}

Considering the moderate values of the Reynolds number in these simulations, a clear inertial range is not established. We  prefer to refer to a restricted  scaling  range (RSR), defined as the range of scales over which different statistics exhibit a discernible scaling, albeit with exponents smaller than those predicted by asymptotic (for infinitely large Reynolds numbers) theories.     

An analytical form of the 4th-order longitudinal structure function can be obtained from Eq. \eqref{eq1} by integrating the sum of  terms 1 to 5:
\begin{equation}
D_{1111}=\frac{1}{r^2} \int_0^r s^2 \left(-Term1-Term{2'}+Term3+Term4+Term5\right) ds.  \label{eq:D4_classic}
\end{equation}
In a statistically steady flow, \textbf{\textit{Term1}} is zero. In the RSR, \textbf{\textit{Term4}} is negligible. One  condition that $D_{1111}$ follows a pure power law is consistent with the requirement  that all terms on the right-hand side of Eq.  \eqref{eq:D4_classic} also exhibit pure power laws, or cancel each other. \textbf{\textit{Term2'}} and \textbf{\textit{Term3}} are shown to follow similar power laws,  while the dissipation source  \textbf{\textit{Term5}} exhibits  a different exponent \citep{Boschung2017}.  It is important to shed some light on the difference between classical turbulence and HVBK quantum turbulence entailed by the mutual friction coupling effect quantified by  \textbf{\textit{Term6}}.

Similar to classical turbulence,  the RSR is not clearly discernible due to the low value of the Reynolds number. Nonetheless, the pressure source \textbf{\textit{Term3}}  perfectly balances transport terms (\textbf{\textit{Term2}} + \textbf{\textit{Term2'}}), while the viscous  \textbf{\textit{Term4}} is negligible.  Unlike the classical turbulence in the RSR, the mutual friction \textbf{\textit{Term6}} acts as  a source term.  Since $D_{111}$ is negative, $\partial_t D_{111} < 0$ reflects vortex stretching enhancement, whilst  $\partial_t D_{111} > 0$ corroborates with reduced vortex stretching. The sign of \textbf{\textit{Term6}} (negative in Eq. \eqref{eq:D3_balance_n} and positive in Eq. \eqref{eq:D3_balance_s}), directly reflects enhanced vortex stretching in the normal fluid (thus, an accelerated cascade) and reduced  vortex stretching and cascade in the superfluid. 

For the normal fluid, \textbf{\textit{Term6}} and the  dissipation source  \textbf{\textit{Term5}} have opposed signs. For decreasing values of the  density ratio $\rho_n/\rho$, the mutual friction \textbf{\textit{Term6}} gradually increases,  which in turn leads to an enhancement  of the dissipation source \textbf{\textit{Term5}}. The physical picture behind this statistical equilibrium between terms  is  that the increase of the vortex stretching rate reflected by  \textbf{\textit{Term5}} requires damping  through the dissipation source term. For the flow to be statistically stationary at the highest normal fluid density ratio ($\rho_n/\rho=0.91$),  only a  small vortex stretching rate has to be introduced by the mutual friction. The normal fluid remains indeed unaffected by the superfluid, thus  behaving  as in classical turbulence. When the superfluid is  dominant ($\rho_n/\rho=0.09$), the mutual friction becomes important in the normal fluid, thus resulting in  a large dissipation source term. At the level of Eqs. \eqref{eq:D4_n} and \eqref{eq:D4_s}, the dissipation source term is non-negligible. This term can effectively modify the scaling exponent of the 4th-order structure functions of velocity increment in the RSR. Interestingly, one can expect that for $0.09<\rho_n/\rho<0.5$, the mutual friction term cancels  the dissipation source term completely. This could trigger an exact $4/3$ scaling exponent for the 4th-order structure functions in the RSR,  for the normal fluid. Therefore, one of our important conclusions is that the normal fluid behaves at very low temperatures as a perfect fluid, since  viscous effects are annihilated by the mutual friction coupling.  

For the superfluid, the mutual friction  \textbf{\textit{Term6}} and the dissipation source \textbf{\textit{Term5}} are  positive and thus  reduce the vortex stretching. When one of them grows, the other one diminishes. In the inviscid limit $\nu_s = 0 $, only \textbf{\textit{Term6}} prevails. When the temperature goes to absolute zero, \textbf{\textit{Term6}} diminishes and  there is no source in the superfluid. In classical turbulence, the scaling exponent of the 4th-order structure functions of the velocity increment in RSR (or in the inertial range) should be $\zeta_4=4/3$ as predicated by the Kolmogorov theory K41. 
In the HVBK model, the viscosity of the superfluid $\nu_s$ is not exactly zero. For large superfluid density ratios ($\rho_s/\rho =0.91$) the mutual friction term is small and the dissipation source term prevails,  being comparable to \textbf{\textit{Term6}}.  For low superfluid density ratios ($\rho_s/\rho =0.09$), the dissipation source term is negligible compared to the mutual friction term in the RSR. Both \textbf{\textit{Term6}} and \textbf{\textit{Term5}} are scale-dependent and they may impact  the scaling exponent of the 4th-order structure functions. 

Finally, the mutual friction terms make the behavior of the 4th-order structure function in the RSR to be more complicated than in classical turbulence. The normal fluid is associated with an enhanced dissipation source term in the RSR for lower and lower temperatures (decreasing $\rho_n/\rho$). In the superfluid, the mutual friction term acts as an addition to the dissipation source term. 


		

		


In the following, we complete our overview of the flow by focusing on the smallest scales, represented by velocity gradients. 


\section{One-point statistics of  velocity gradients} \label{section_onepoint}

\begin{figure}[!h]
\centering
  \begin{subfigure}{0.45\textwidth}
   		\includegraphics[width=1.\textwidth,height=0.7\textheight,keepaspectratio,trim={1.2cm 7.2cm 1.4cm 8.1cm}, clip]{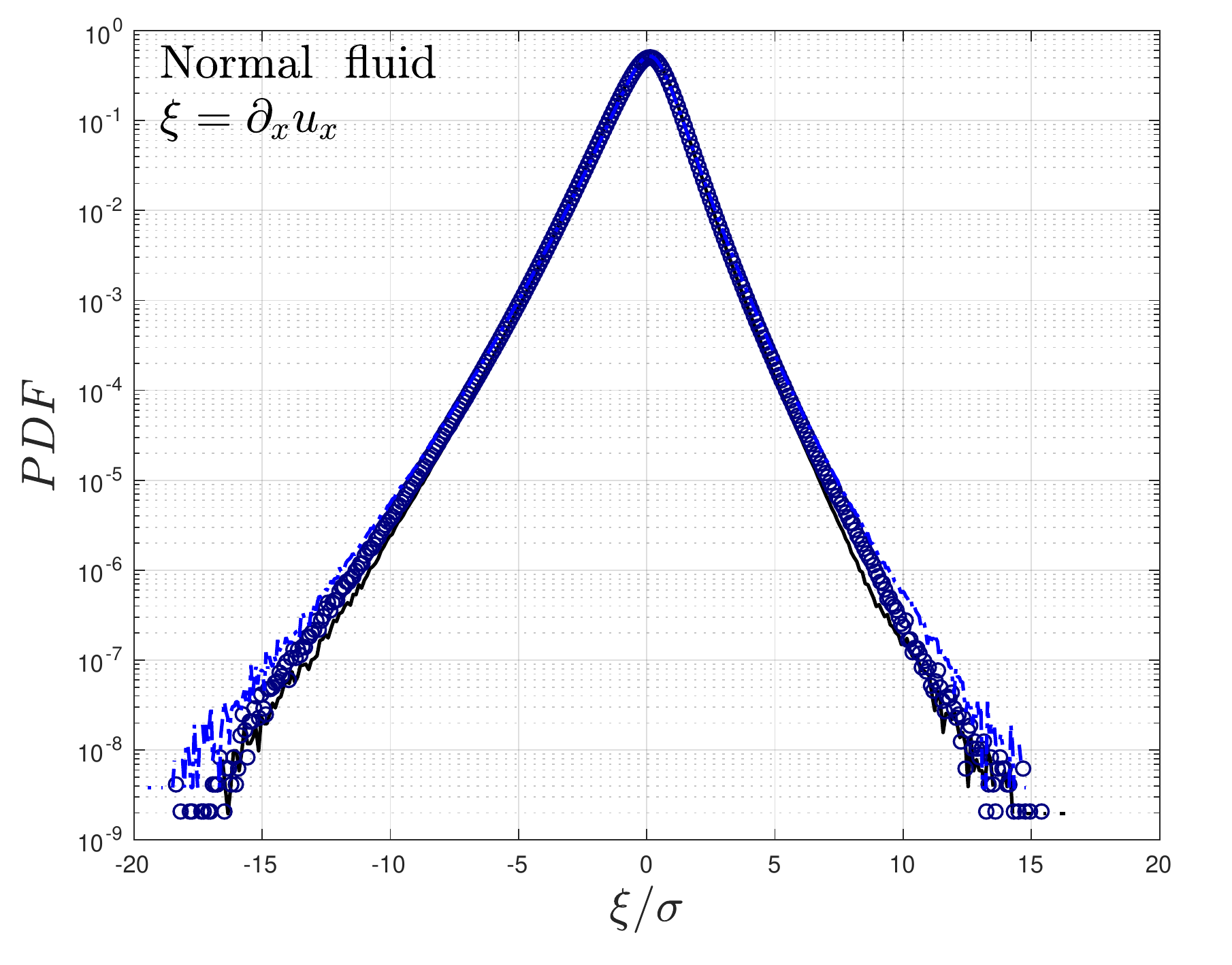}
  \caption{ }
  \label{fig:grad-pdf-normal}
  \end{subfigure}%
  \begin{subfigure}{0.45\textwidth}
		\includegraphics[width=1.\textwidth,height=0.7\textheight,keepaspectratio,trim={1.2cm 7.2cm 1.4cm 8.1cm}, clip]{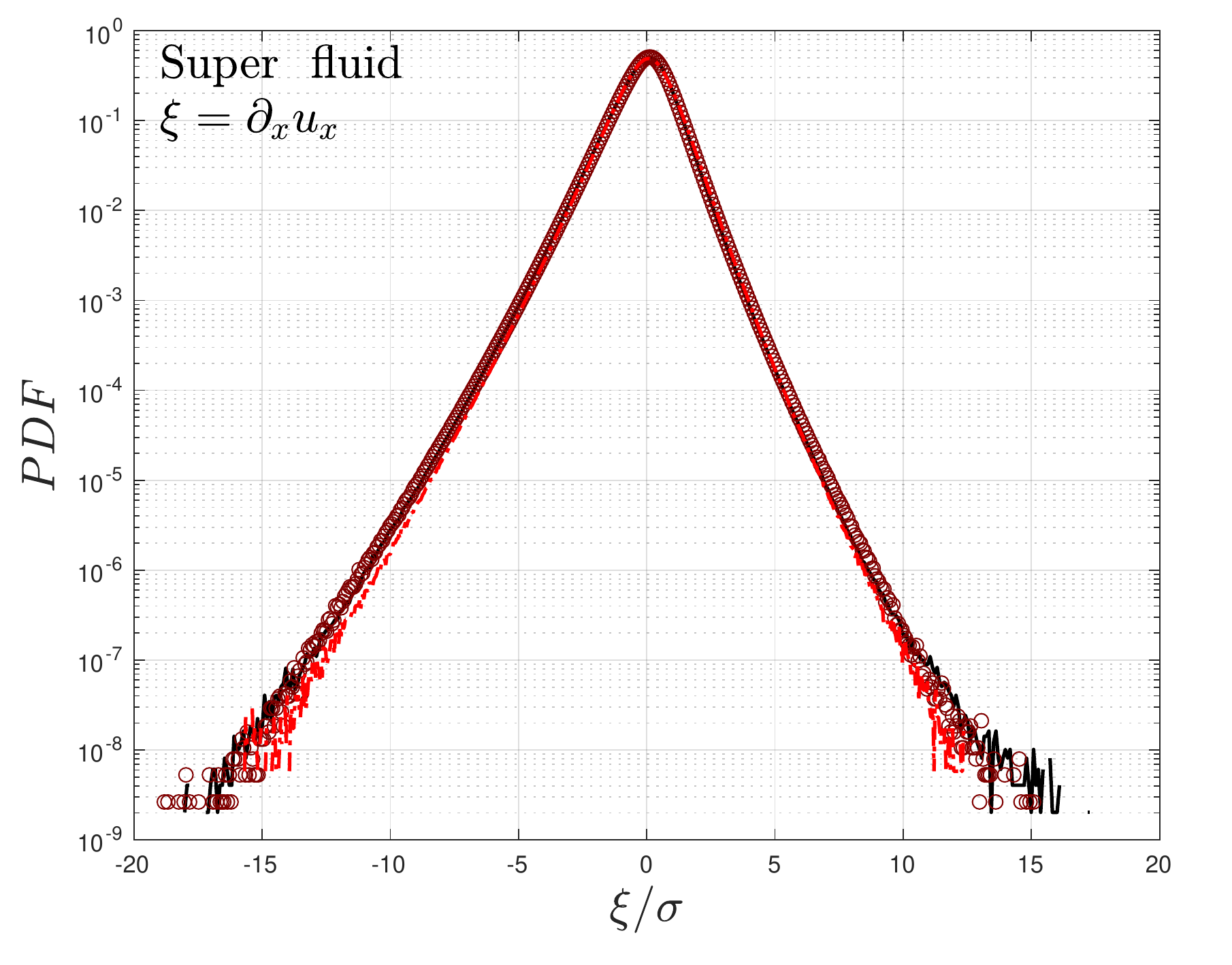}
  \caption{  }
  \label{fig:grad-pdf-super}
  \end{subfigure}%
\hfill
    \begin{subfigure}{0.45\textwidth}
   		\includegraphics[width=1.\textwidth,height=0.7\textheight,keepaspectratio,trim={1.2cm 7.2cm 1.4cm 8.1cm}, clip]{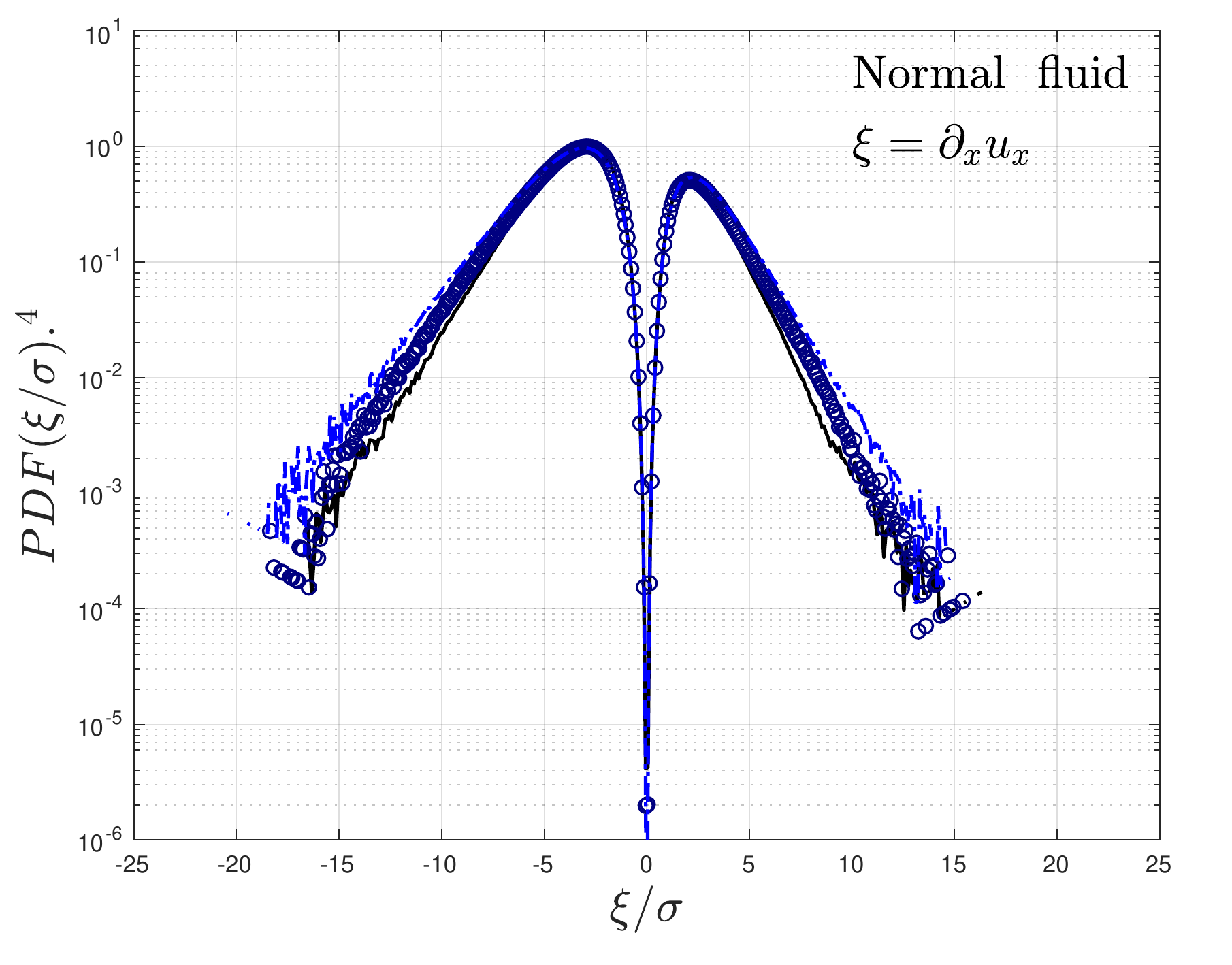}
  \caption{ }
  \label{fig:grad-pdf-normalX4}
  \end{subfigure}%
  \begin{subfigure}{0.45\textwidth}
		\includegraphics[width=1.\textwidth,height=0.7\textheight,keepaspectratio,trim={1.2cm 7.2cm 1.4cm 8.1cm}, clip]{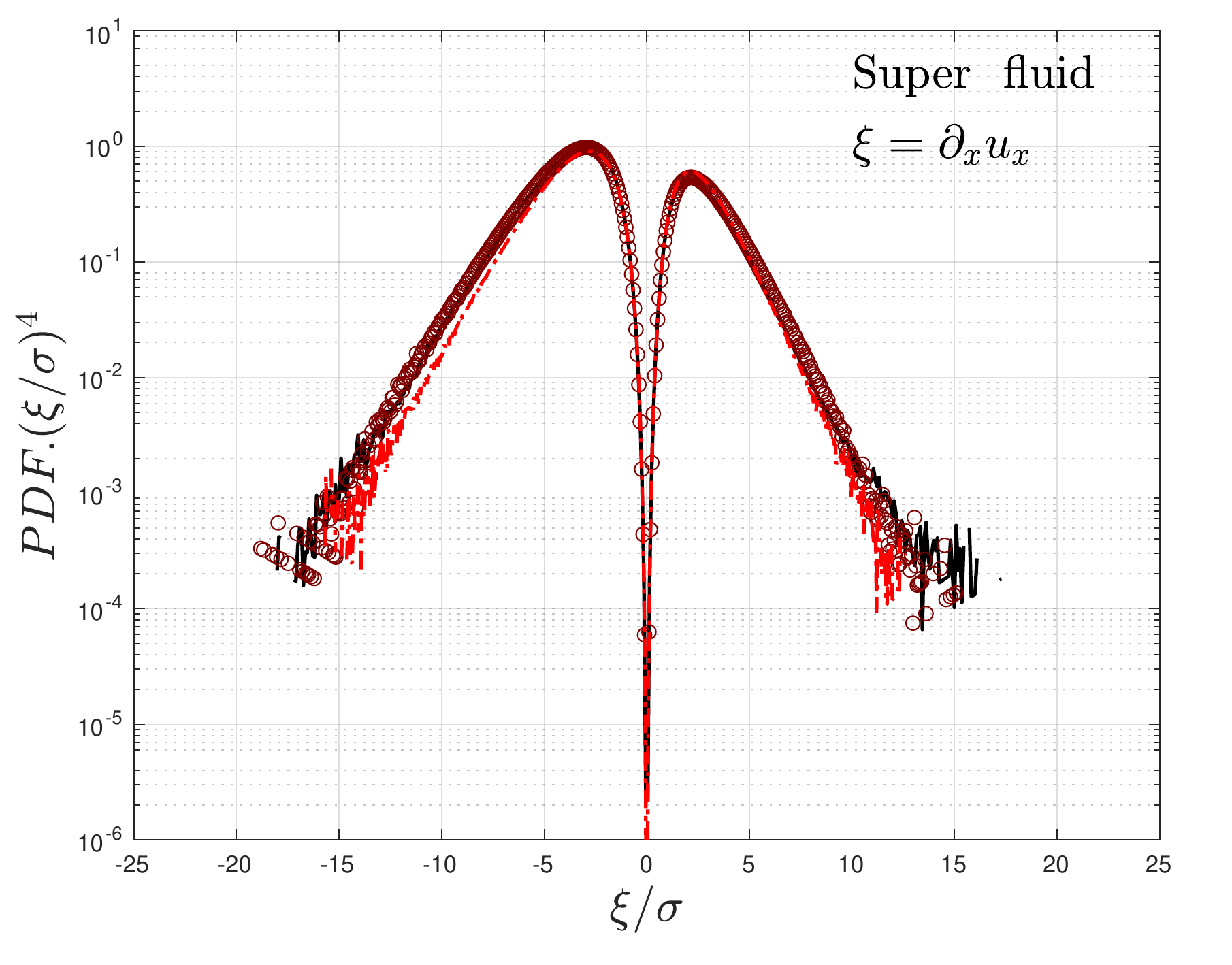}
  \caption{ }
  \label{fig:grad-pdf-superX4}
  \end{subfigure}%
  \caption{PDFs of gradients of longitudinal velocity in the normal fluid (a) and superfluid  (b). Panels (c) and (d) show corresponding
  	normalized PDFs  as $(\xi/\sigma)^4PDF$, with $\sigma = \sqrt{\langle \xi^2 \rangle - \langle \xi \rangle^2}$ the standard deviation of the velocity gradient. Results for three density ratios: (- $\cdot$) $\rho_n/\rho =0.09$, ($\circ$) $\rho_n/\rho =0.5$, (--) $\rho_n/\rho =0.91$.}
  \label{fig:grad-pdf}
\end{figure}

We focus on one-point statistics of the small-scale motion. Particular emphasis is put on the flatness of the velocity gradient, which reflects the effect of turbulence intermittency on small scales dynamics. The probability density function (PDF) of the longitudinal velocity gradient $\xi = \partial_x u$, for the same  density ratios as previously ($\rho_n/\rho=0.91,\ 0.5,\ 0.09$), are shown in Figs. \ref{fig:grad-pdf-normal} and \ref{fig:grad-pdf-super}  for the normal and superfluid components, respectively. Similar to classical turbulence, PDFs exhibit non-Gaussian skewed shapes, with stretched tails skewed towards negative values of the velocity gradients. Note that negative values of velocity gradients are much larger than its variance.  For decreasing values of  the normal fluid density ratios, PDFs tails become more and more stretched. However, PDFs of the velocity gradients in the superfluid show non-monotonic trends. 

The integration over the whole domain leads  to the $p$th-order moment of $\xi$:
\begin{equation}
    \langle \xi^p \rangle = \int_{-\infty}^{\infty}\xi^p PDF(\xi) d\xi.
    \label{eq:moments}
\end{equation}
The normalized 4th-order moment is the flatness factor:
\begin{equation}
    F = \frac{\langle \xi^4\rangle}{\langle \xi^2\rangle ^2}.
    \label{eq:flatness}
\end{equation}

\begin{figure}[!h]
	\centering
	\includegraphics[width=.5\textwidth,height=0.5\textheight,keepaspectratio,trim={1.5cm 7.2cm 2.4cm 7.5cm}, clip]{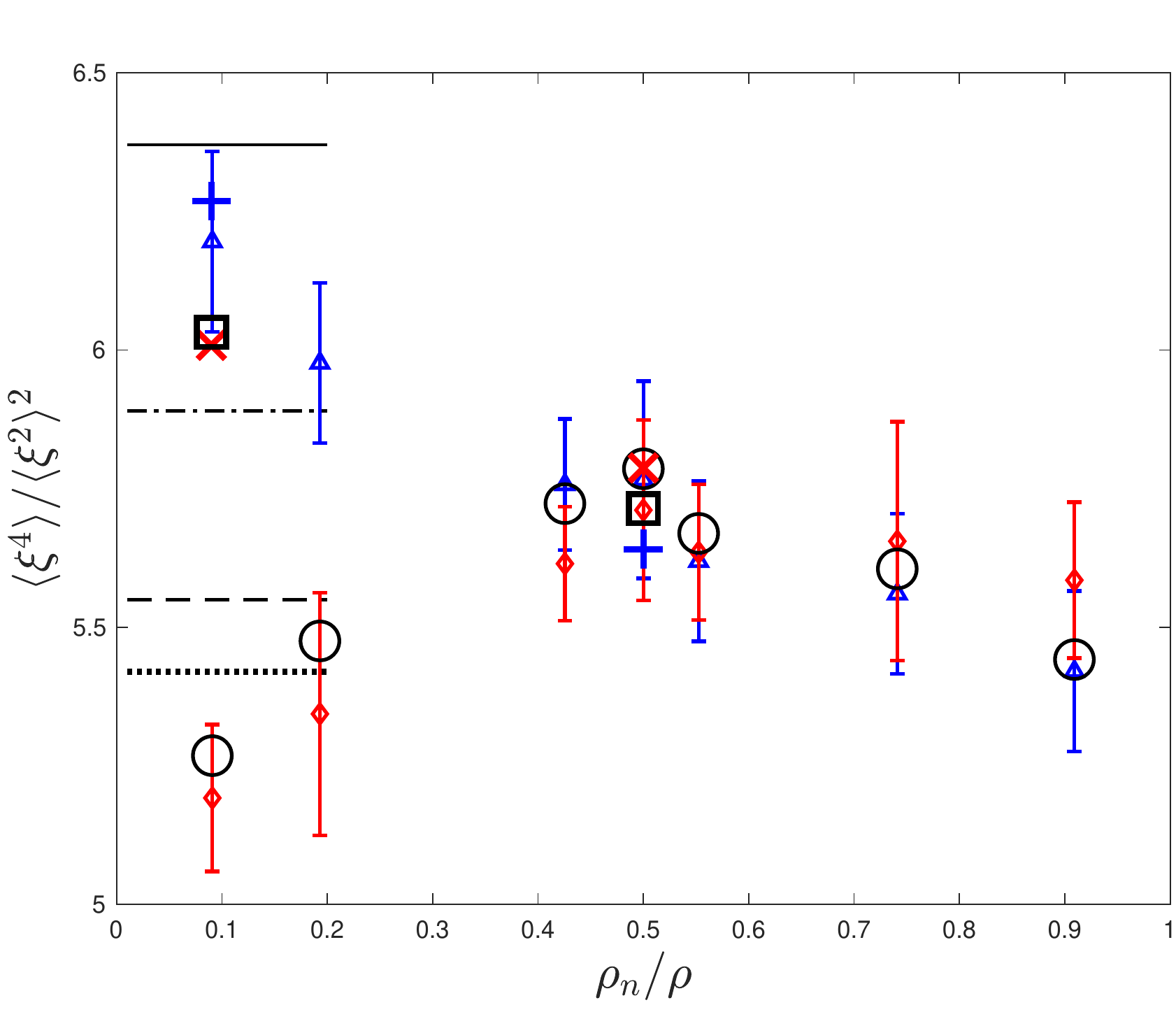}
	\caption{Flatness factors of the longitudinal velocity gradient $\xi =\partial_x u$ versus density ratio $\rho_n/\rho$ for the normal fluid ($\triangle$), the superfluid ($\diamond$)  and total fluid ($\bigcirc$). Error bars are the root-mean-square value of the variance of the flatness factors computed with 20 to 50 snapshots, and $10^8$ data points for each snapshot.  Horizontal lines mark the flatness factor 
		computed from DNS of classical turbulence  (\cite{Ishihara_2007}): ($\cdot \ \cdot$) $Re_\lambda =94.6$, (-  -) $R_\lambda =94.4$, (- $\cdot$) $R_\lambda =167$. (--) $R_\lambda =173$.
		All points are computed for $N=512$, except the following ones, based on  $N=1024$ resolution: big blue $+$ (normal fluid), big red $\times$ (superfluid) and big black $\square$ (total fluid). }
	\label{fig:S4:derivative}
\end{figure}

Figure \ref{fig:grad-pdf} shows PDFs of gradients of longitudinal velocity (panels a and b) and normalized PDFs (panels c and d) as 
 $(\xi/\sigma)^4PDF$, where $\sigma = \sqrt{\langle \xi^2 \rangle - \langle \xi \rangle^2}$ is the standard deviation of the velocity gradient.
 PDFs are well converged for large events, with errors smaller than $1\%$. This signifies  that flatness factors computed from the PDFs are accurate. Flatness factors  are plotted in Fig.  \ref{fig:S4:derivative} for all considered cases. For the normal fluid, the flatness factor increases monotonically when $\rho_n/\rho$ diminishes, which indicates that the intermittency in normal fluid is enhanced for lower and lower temperatures.  The superfluid follows the same trend as the normal fluid. 

This observation can be explained by the energy exchange between the two fluid components. On average, mutual friction acts as a  source term that enhances energy at all scales in the normal fluid. Since energy input is expected to occur mainly at locations with strong vorticity, we suggest the following scenario.
First, vorticity distributions in the two fluids are coherent (aligned) and the mutual friction depends directly on the magnitude of the vorticity.
As the relative velocity seems to be uniformly distributed in space,  with a Gaussian PDF, we infer  that the mutual friction enhances locally the vorticity and thus the intermittency. When the density ratio $\rho_n/\rho$ decreases, the mutual friction term in the normal fluid momentum equation is more important. As a consequence, the intermittency grows when  the temperature diminishes.    The superfluid is strongly locked with the normal fluid, thus following a similar trend.  

We further compute the flatness of the total longitudinal velocity gradient  $\partial_x u = \frac{\rho_n}{\rho}\partial_x u_n +\frac{\rho_s}{\rho}\partial_x u_s$, with $u_n$ and $u_s$ the longitudinal velocity components in the normal fluid and superfluid, respectively.
Flatness factors of  the total longitudinal velocity gradient are plotted against $\rho_n/\rho$ in Fig. \ref{fig:S4:derivative}. The flatness factor is controlled by the normal fluid for high $\rho_n/\rho$, and by the superfluid for low $\rho_n/\rho$.  The intermittency of the total fluid  continuously increases when the temperature diminishes. 

In Fig. 
\ref{fig:S4:derivative}, horizontal short lines indicate DNS results of classical turbulence flatness factor for different values of $R_\lambda$ \citep{Ishihara_2007}.  For Reynolds numbers close to that considered in our simulations ($R_\lambda \approx 94$),  the flatness $F$ in classical turbulence ranges between $5.42$ and $5.55$. These values are very close to the flatness  we obtained  for $\rho_n/\rho = 0.91$, corresponding to $R_\lambda \approx 90$ in the normal fluid. When $\rho_n/\rho$ decreases,   $R_\lambda$ also decreases in normal fluid (due to the mutual friction), and the flatness factor increases for the total fluid. We obtained the value  $F=5.786$ for $\rho_n/\rho=0.5$. While a resolution of $N=512$ leads to a flatness that drops back to $F=5.268$ for $\rho_n/\rho=0.09$, an enhanced resolution of $N=1024$ leads to values comparable for both normal and superfluid, thus emphasizing their locking.

HVBK quantum turbulence simulated here exhibits  the same degree of intermittency  as observed in classical turbulence.

\section{The flatness of the velocity derivative in superfluid turbulence} \label{section_flatnessderivative}

At this stage, it is important to go back to the theoretical framework provided by  first principles (here the HVBK equations) and to consider the limiting behavior of Eqs. \eqref{eq:D3_balance_n} and \eqref{eq:D3_balance_s}.


To obtain the expression of the flatness derivative, we consider that $r \rightarrow 0$ and apply a Taylor series expansion up to the fifth-order in $r$ \citep{Shunlin2017,Shunlin2018}.  Using the homogeneity hypothesis, we obtain:   
\begin{equation}
	\frac{\partial }{{\partial x}} \moy{{{\left( {\frac{{\partial u}}{{\partial x}}} \right)}^2}{{\left( {\frac{{{\partial ^2}u}}{{\partial {x^2}}}} \right)}}}   = 0{\rm{  }} \Longrightarrow {\rm{  }} 2  \moy{\left(\frac{\partial u}{\partial x}\right) \left( \frac{\partial^2 u}{\partial x^2}\right)} =  -\moy{\left(\frac{\partial u}{\partial x}\right)^2 \left( \frac{\partial^3 u}{\partial x^3}\right)}, 	
\end{equation}
and hence
\begin{equation}
	\moy{(\delta u)^3} \simeq \moy{\left(\frac{\partial u}{\partial x}  \right)^3} r^3 {\color{black}- \frac{1}{4}} \moy{\left(\frac{\partial u}{\partial x}\right)  \left( \frac{\partial^2 u}{\partial x^2}\right)^2}  r^5+.... \label{eq4} 
\end{equation}
The 4th-order structure function can be written as  
\begin{equation}
	D_{1111}= \moy{(\delta u)^4} \simeq \moy{
		\left( \frac{\partial u}{\partial x}\right)^4} r^4+ ... \label{eq5} 
\end{equation}
and similarly
\begin{equation}
	D_{1122}= \langle{\left(\delta u \right)^2\left(\delta v \right)^2} \rangle \simeq \moy{
		\left( \frac{\partial u}{\partial x}\right)^2 \left( \frac{\partial v}{\partial x}\right)^2} r^4+....  \label{eq6}
\end{equation}

An equation for $F$, the velocity derivative flatness factor \eqref{eq:flatness}, can be obtained \citep{Shunlin2018}  by applying the following operator ${\cal {O}}$  to terms in Eqs.  \eqref{eq:D3_balance_n} and  \eqref{eq:D3_balance_s}:
\begin{equation}
	{\cal {O}}(Term) \equiv  \lim_{r\rightarrow 0} \frac{\frac{Term}{r^3}}{\frac{\moy{(\delta u)^2}^2}{r^4}}=\lim_{r\rightarrow 0} 
	r \cdot \frac{Term}{\moy{(\delta u)^2}^2}. 
\end{equation}

We obtain that 
\begin{equation}
	{\cal O} (Term 2')= -6 \moy{
	\left( \frac{\partial u}{\partial x}\right)^2 \left( \frac{\partial v}{\partial x}\right)^2}  /  \moy{
	\left( \frac{\partial u}{\partial x}\right)^2}^2 = -6 S_{uv,2},
\end{equation}
where the notation $ S_{uv,2}$ was introduced for the sake of simplicity. 
The pressure term becomes, once $\cal O$  is applied:  
\begin{equation}
	{\cal O} (Term 3) = -\frac{\ds 3   \moy{\left(\frac{\partial u}{\partial x}\right)^2 \frac{\partial^2 p }{\partial x^2} }}{\ds \moy{\left(\frac{\partial u}{\partial x}\right)^2}^2}.
\end{equation}
After applying the operator $\cal O$ to the coefficient  $r^3$, the dissipation term leads to 
\begin{equation}
	{\cal O} (Term 4)|_{r^3} = \frac{\ds 9\nu \moy{ \left( \frac{\partial^3 u}{\partial x^3}\right)\left( \frac{\partial u }{\partial x}\right)^2 } }{\ds \moy{\left(\frac{\partial u}{\partial x}\right)^2}^2}.
\end{equation}

\textbf{\textit{Term5}}  leads to a linear combination
\begin{equation}
	{\cal O} (Term 5)|_{r^3} = -\frac{\ds 2\nu \moy{ \left( \frac{\partial^3 u}{\partial x^3}\right)\left(\frac{\partial u }{\partial y}\right)^2 } +4\nu \moy{ \left(\frac{\partial^3 u}{\partial x^3}\right)\left(\frac{\partial u }{\partial x}\right)^2}}{\ds \moy{\left(\frac{\partial u}{\partial x}\right)^2}^2}.
\end{equation}

\textbf{\textit{Term6}}, specific to HVBK equations, leads after Taylor series expansion and application of the operator $\cal O$, for the normal fluid:
\begin{equation}
	{\cal O} (Term 6)^n \sim \frac{\ds 3 \frac{\rho_s}{\rho} \moy{\left(\frac{\partial u}{\partial x}\right)^2\left(\frac{\partial F_{\parallel}}{\partial x}\right)^{ns}}}{\ds{\moy{\left(\frac{\partial u}{\partial x}\right)^2}^2}},  \label{eq_term6_normal}
\end{equation}
and for the superfluid
\begin{equation}
	{\cal O} (Term 6)^s \sim -\frac{\ds 3 \frac{\rho_n}{\rho} \moy{\left(\frac{\partial u}{\partial x}\right)^2\left(\frac{\partial F_{\parallel}}{\partial x}\right)^{ns}}}{\ds{\moy{\left(\frac{\partial u}{\partial x}\right)^2}^2}}.  \label{eq_term6_superfluid}
\end{equation}
Similarly, \textbf{\textit{Term7}} leads for the normal fluid
\begin{equation}
	{\cal O} (Term 7)^n \sim \frac{\ds 3  \moy{\left(\frac{\partial u}{\partial x}\right)^2\left(\frac{\partial f_{\parallel}}{\partial x}\right)^{n}}}{\ds {\moy{\left(\frac{\partial u}{\partial x}\right)^2}^2}},  \label{eq_term7_normal}
\end{equation}
and for the superfluid
\begin{equation}
	{\cal O} (Term 7)^s \sim \frac{\ds 3  \moy{\left(\frac{\partial u}{\partial x}\right)^2\left(\frac{\partial f_{\parallel}}{\partial x}\right)^{s}}}{\ds{\moy{\left(\frac{\partial u}{\partial x}\right)^2}^2}}.   \label{eq_term7_superfluid}
\end{equation}

The limiting form of Eqs. \eqref{eq:D3_balance_n} and \eqref{eq:D3_balance_s} as $r \rightarrow 0$ can be finally presented as
\begin{align}
		\label{eq_flat_general} 
	{6(F^n-S_{uv,2}^n)}  &=  {\cal O} (Term 3)^n +{\cal O} (Term 4)^n|_{r^3} +{\cal O} (Term 5)^n|_{r^3}+ {\cal O} (Term 6)^n+{\cal O} (Term 7)^n,\\
	{6(1-S_{uv,2}^s/S_4^s) F^s}  &=  {\cal O} (Term 3)^s +{\cal O} (Term 4)^s|_{r^3} +{\cal O} (Term 5)^s|_{r^3}+ {\cal O} (Term 6)^s+{\cal O} (Term 7)^s,
	\label{eq_flat_general_final} 
\end{align}

\cite{Lyazid2017} showed that $S_{uv,2} / F \approx 0.85$ if $Re_\lambda >200$ and this constant is independent of the Reynolds number. For the present study, values of  $S_{uv,2} / S_4$ are shown for different density ratios in Tab. \ref{table_Oterms}. These values remain almost unchanged for the $\rho_n/\rho = 0.91, \ 0.5 $, but for $\rho_n/\rho = 0.09$, $S_{uv,2} / S_4$ slightly diminishes and drops to $0.75$ for the normal fluid, and $0.705$ for the superfluid. 
\begin{table}
	\centering
	\begin{tabular}{c |c c c c c}
		\hline
		\ & $\rho_n/\rho$ &$\langle{(\delta u)^2}\rangle^2$&$\langle{(\delta u)^4}\rangle $&$\langle(\delta u)^2(\delta v)^2 \rangle$&$\langle{(\delta u)^2(\delta v)^2}\rangle /  \langle{(\delta u)^4}\rangle $\\
		\hline
		n&0.91& 6.3314e-10 & 3.2965e-09 & 2.7139e-09 & 0.8233\\
		n&0.50& 1.3595e-09 & 7.5912e-09 & 6.2411e-09 & 0.8221\\
		n&0.09& 4.7415e-09 & 2.8769e-08 & 2.1679e-08 & 0.7536\\
		
		\hline
		
		s&0.91&   1.2436e-09 & 6.6820e-09 & 5.4091e-09 & 0.8095\\
		s&0.50&   3.5142e-09 & 1.9483e-08 & 1.5659e-08 & 0.8037\\
		s&0.09&   2.3925e-08 & 1.2353e-07 & 8.7167e-08 & 0.7056\\
		\hline
	\end{tabular}
	\caption{ Values used in expressing the flatness factor at the smallest scales in the limit  $r\rightarrow0$. Practically $(r\rightarrow \Delta$), with $\Delta$ the smallest grid size in simulations (see Tab.  \ref{table_param-II} for the Kolmogorov normalized mesh size).}
	\label{table_Oterms}
\end{table}

Table \ref{table_Oterms_I} shows that all terms in Eq. \eqref{eq_flat_general_final} are very well balanced. This proves that all terms are correctly accounted for.  
 In the normal fluid, the balance between different terms is reached within an error of   $0.05 \%$ for the flatness $F$.  ${\cal O}(Term3)$ increases as the temperature diminishes (the normal fluid is less and less present). Viscous terms are not negligible in the case of present $Re_\lambda$. The combined contribution  of ${\cal O}(Term4)$+${\cal O} (Term5)$ increases, but this enhancement is counter-balanced  by the mutual friction force contribution ${\cal O} (Term 6)$.  The external force was neglected, as usually done for larger scales in classical turbulence. 

\begin{table}
	\centering
	\begin{tabular}{c | c c c c c c c c }
		\hline
		
		\ &$\rho_n/\rho$ &${\cal O} (Term 2)$ &${\cal O} (Term 2')$ & ${\cal O} (Term 3)$ & ${\cal O} (Term 4)$ & ${\cal O} (Term 5)$  & ${\cal O} (Term 6)$  & ${\cal O} (Term 7)$ \\
		\hline
		
		n&0.91 & 29.6833  & -25.7186 & 4.2801  & 2.5938  & -1.3144 &  -0.0092   &  -0.0074 \\
		n& 0.5 & 32.8070  & -27.5455 & 4.4541  & 3.7629  & -1.8098 &  -0.8888   &  -0.0046 \\
		n&0.09 & 35.9923  & -27.4329 & 6.5976  & 5.7996  & -2.2460 &  -3.1359   &  -0.0020 \\
		
		\hline
		s&0.91 & 31.8223  & -26.0972 & 4.8648  & 0.4632  & -0.1609    & 0.9296   &  -0.0047\\
		s&0.5. & 34.4168  & -26.7352 & 4.8345  & 0.5350  & -0.1764    & 0.6823   &  -0.0023\\
		s&0.09 & 33.2927  & -21.8602 & 6.3783  & 0.4023  & -0.1070    & 0.1321   &  -0.0006\\
		\hline
	\end{tabular}
	\caption{DNS results for terms resulting after applying  the operator to ${\cal O}$ in Eq. \eqref{eq_flat_general_final}, for the normal fluid (first three lines) and the superfluid (last three lines)). }
	\label{table_Oterms_I}
\end{table}

For the superfluid, only the pressure-related  term ${\cal O}(Term 3)$ matters. This result corroborates with a scenario valid for large Reynolds numbers in classical turbulence.   ${\cal O}(Term 3)$ increases monotonically when  the temperature decreases. For the lowest temperature, the ratio $S_{uv,2} /F$ is the smallest and  both $S_{uv,2}$ and $F$ are smaller than for higher temperatures.   This behaviour is corroborated with    the spectral cut-off inherently introduced in DNS at the inter-vortex scale, which  leads to an under estimation of high-order moments of small scales (here, represented by velocity  gradients). However, this behaviour can have a physical explanation in the superfluid helium by the energy accumulation at scales close to the inter-vortex scale. 
We finally note that for $\rho_n/\rho=0.09$ the terms are not well balanced as in other cases and errors are up to  $15\%$. This is due to the limited resolution for the superfluid at low temperatures.

\section{Conclusion} \label{section_conclusions}

We used direct numerical simulations of the HVBK model to inspect, for different density ratios,  the behaviour of the 4th-order structure function, as resulting from the transport equation of the 3rd--order structure function. Starting from the HVBK equations  for two fluids, we derived the 3rd-order structure function transport equations in both normal and superfluid. 
Within the Restricted Scaling Range, we found that the mutual friction does not modify significantly the dynamics  of viscous scales. Similar to the classical turbulence, viscous terms and  dissipative source terms are less important than the other terms. The mutual friction term acts differently for the two fluid components. For the normal fluid, the mutual friction term has an opposite sign with respect to the dissipation source terms.  Depending on the density ratios,   it can diminish, and even completely cancel the action of  the dissipation source term.  
For the lowest temperature, we show that the normal fluid behaves, in the RSR,  as a fluid with zero viscosity.  
In superfluid, the mutual friction term is mostly irrelevant. It can be neglected when compared to the transport terms and the pressure source term. The dissipation source term, introduced by the (artificial) superfluid viscosity, has the same sign as the mutual friction term, finally resulting in a diminished cascade and reduced small-scales intermittency,  as reflected by the flatness of the velocity derivative. 
Note that RSR intermittency effects are not addressed here, as the conventional assessment of the scaling exponents is limited by the relatively low values of the Reynolds numbers. 

We also used one-point statistics, PDFs of longitudinal velocity gradient to analyse the temperature dependency of small-scale  intermittency of quantum turbulence. We conclude that both the normal fluid and superfluid intermittency is enhanced when $\rho_n/\rho$  is decreasing.   This is consistent with the strong locking of the two fluids. The flatness factors are also found in reasonable agreement with classical turbulence. 
 Further perspectives of this work include the account of a more general expression of the friction force, based upon at least one additional equation for e.g. the vortex  line density \citep{nemirovskii2020}. Another open question is the  coupling between Navier-Stokes like equations with Gross-Pitaevskii equation for very small scales and very low temperatures.



\vspace{1cm}

The authors declare no conflict of interest. 

\section{Acknowledgements}

The authors acknowledge financial support from the French ANR grant ANR-18-CE46-0013 QUTE-HPC. This work used computational resources provided by  CRIANN (Centre R{\'e}gional Informatique et d'Applications Num{\'e}riques de Normandie).
Drs. M. Gauding and G. Sadaka are warmly thanked for technical support and useful discussions.


\section{Appendix 1. Robustness of the results for a smaller viscosity ratio} \label{appendix_viscratios}

\begin{figure}[!h]
	\centering
	\includegraphics[width=0.45\textwidth,height=0.35\textheight,keepaspectratio,trim={1.5cm 7.2cm 2.4cm 8.5cm}, clip]{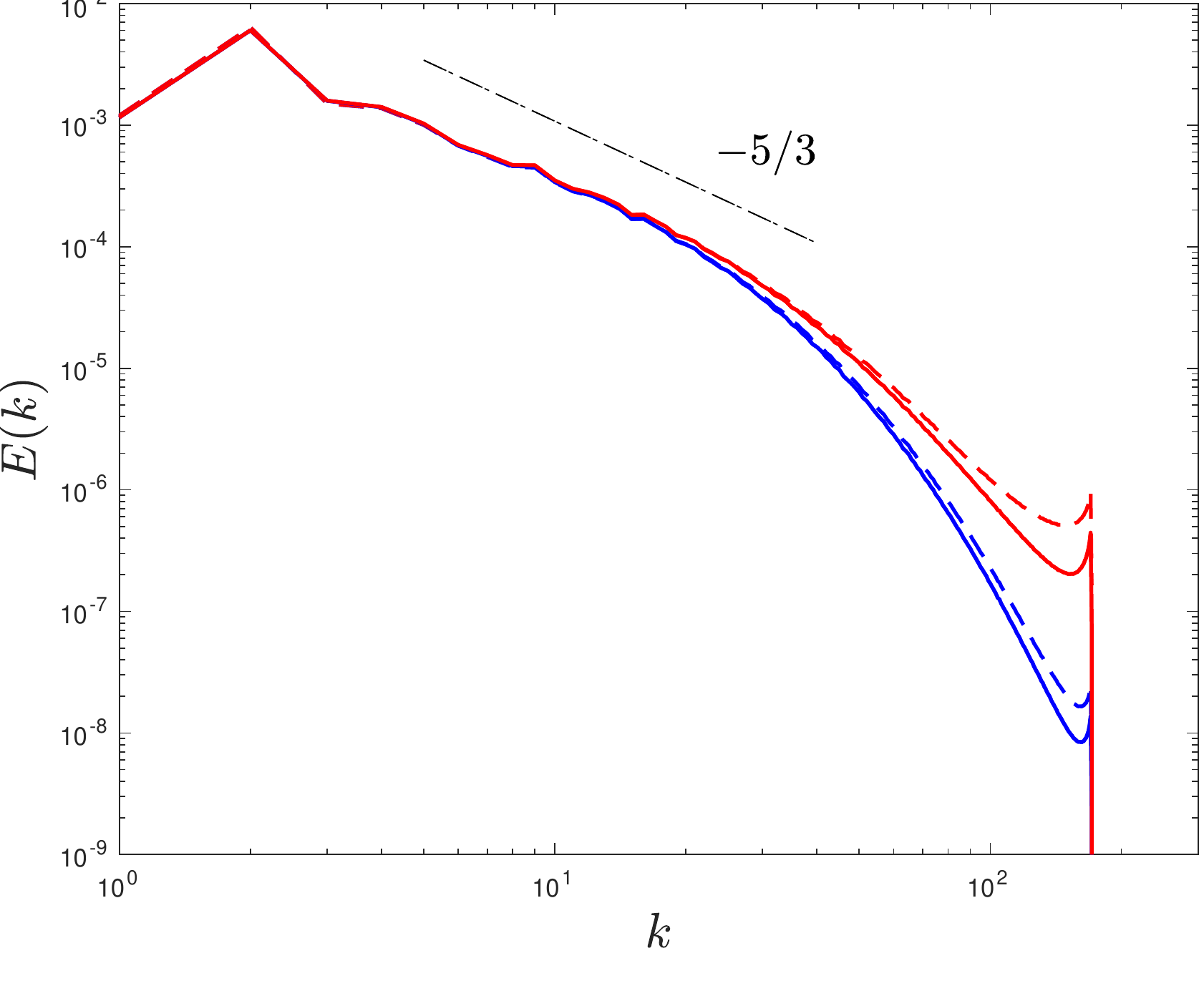}
	\caption{Energy spectrum for $\rho_s/\rho_n=1$ and different viscosity ratios; (-\ -) for $\nu_s/\nu_n = 0.025$ and ($-$) for $\nu_s/\nu_n = 0.1$. ($-\, \cdot$) marks the power law of $-5/3$. }
	\label{fig:Ek_rho1}
\end{figure}

	We have performed numerical simulations for a different viscosity ratio, i.e. $0.025$, a quarter of that initially studied and reported in the corpus of the paper.  
Figure \ref{fig:Ek_rho1} shows the energy spectrum for different viscosity ratios, and  for the density ratio of $\rho_s/\rho_n = 1$. For smaller viscosity ratio $\nu_s/\nu_n=0.025$, the energy content at the level of the cut-off scales is slightly larger than for the  viscosity ratio $\nu_s/\nu_n=0.1$. While this  result is obvious, as the superfluid dissipates less, the difference is negligible. The dissipation scale $\eta_s = (\nu_s^3/\varepsilon_s)^{1/4}$ (the Kolmogorov scale) for the smaller viscosity ratio $\nu_s/\nu_n=0.025$ is significantly reduced, and equal to   $\eta_s = 0.0017$, whereas it was of  $\eta_s = 0.0034$ for $\nu_s/\nu_n=0.1$. 
For $\nu_s/\nu_n=0.025$, the viscosity of the superfluid decreases, while  the mean energy dissipation rate of the superfluid increases. The dissipation rate for $\nu_s/\nu_n=0.1$ was of $\varepsilon_s = 1.8e-4$, while  for $\nu_s/\nu_n=0.025$ we compute  $\varepsilon_s = 5.5463e-4$. The reason of this increase is the accumulation of the energy at small scales, resulting in an increase of velocity gradients. 

\begin{figure}[!h]
\begin{subfigure}{0.5\textwidth}
	\includegraphics[width=1.\textwidth,height=0.7\textheight,keepaspectratio,trim={1.5cm 7.2cm 2.4cm 7.5cm}, clip]{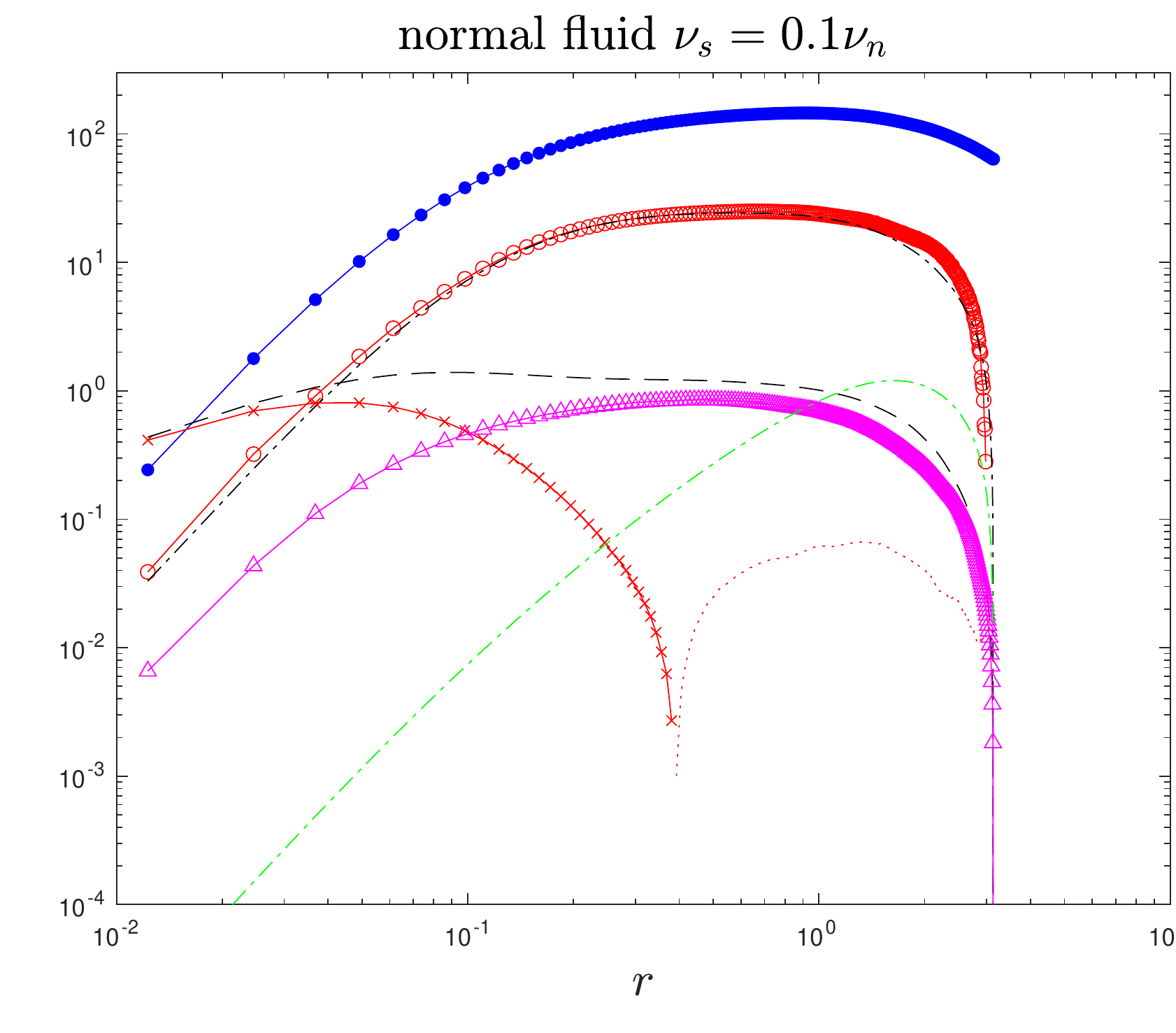}
\end{subfigure}%
\begin{subfigure}{0.5\textwidth}
	\includegraphics[width=1.\textwidth,height=0.7\textheight,keepaspectratio,trim={1.5cm 7.2cm 2.4cm 7.5cm}, clip]{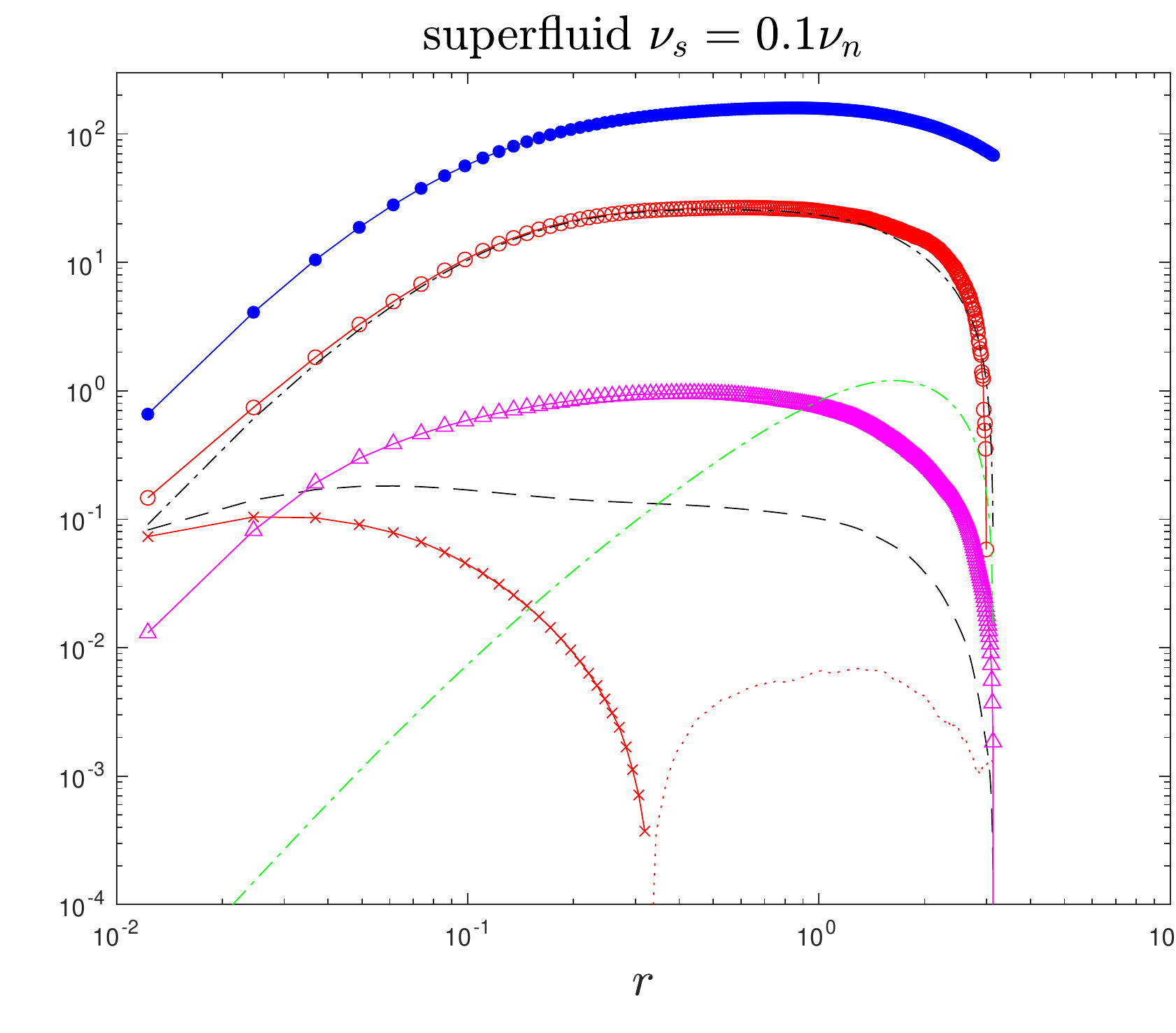}
\end{subfigure}%
\hfill
\begin{subfigure}{0.5\textwidth}
	\includegraphics[width=1.\textwidth,height=0.7\textheight,keepaspectratio,trim={1.5cm 7.2cm 2.4cm 7.5cm}, clip]{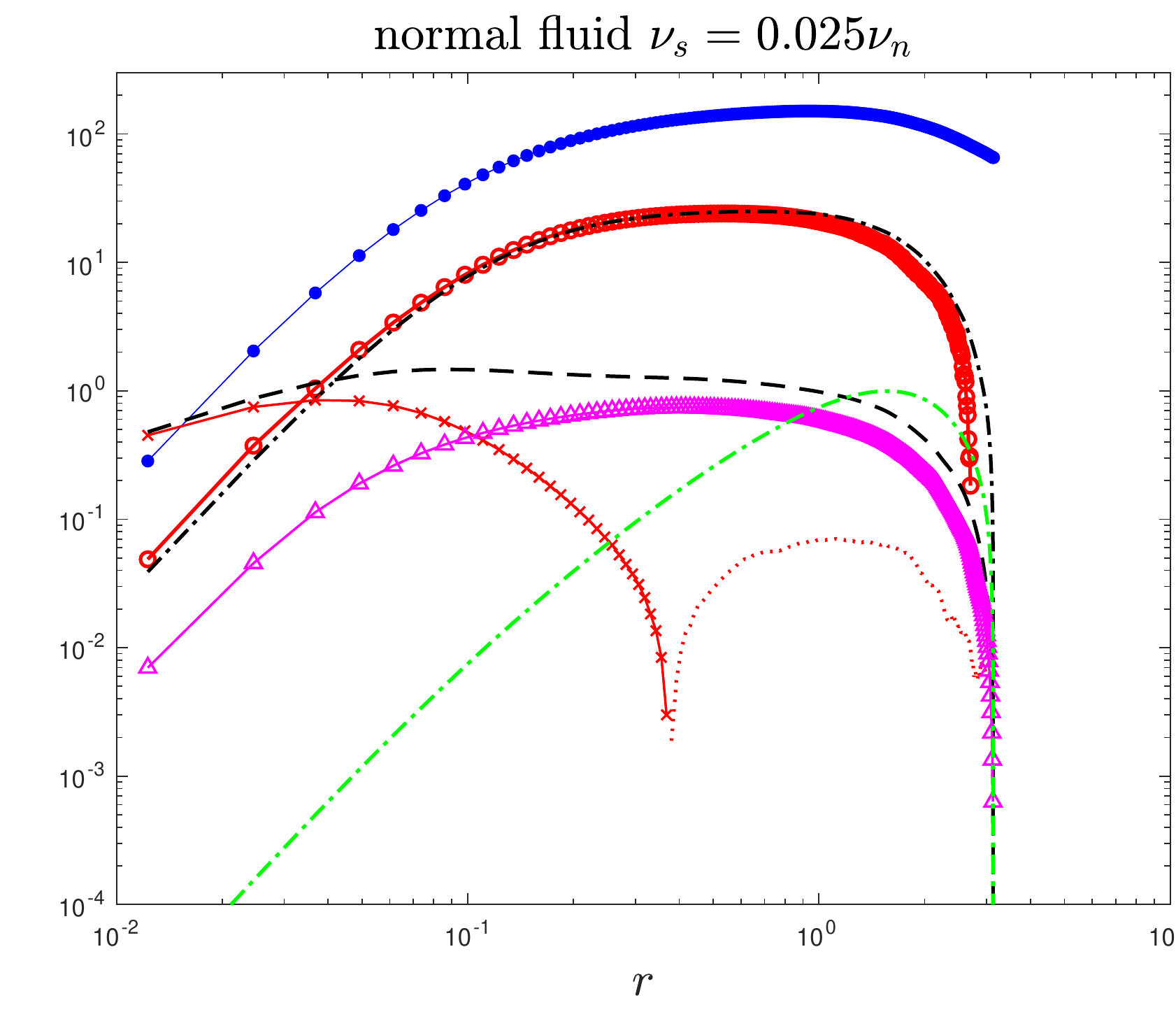}
\end{subfigure}%
\begin{subfigure}{0.5\textwidth}
	\includegraphics[width=1.\textwidth,height=0.7\textheight,keepaspectratio,trim={1.5cm 7.2cm 2.4cm 7.5cm}, clip]{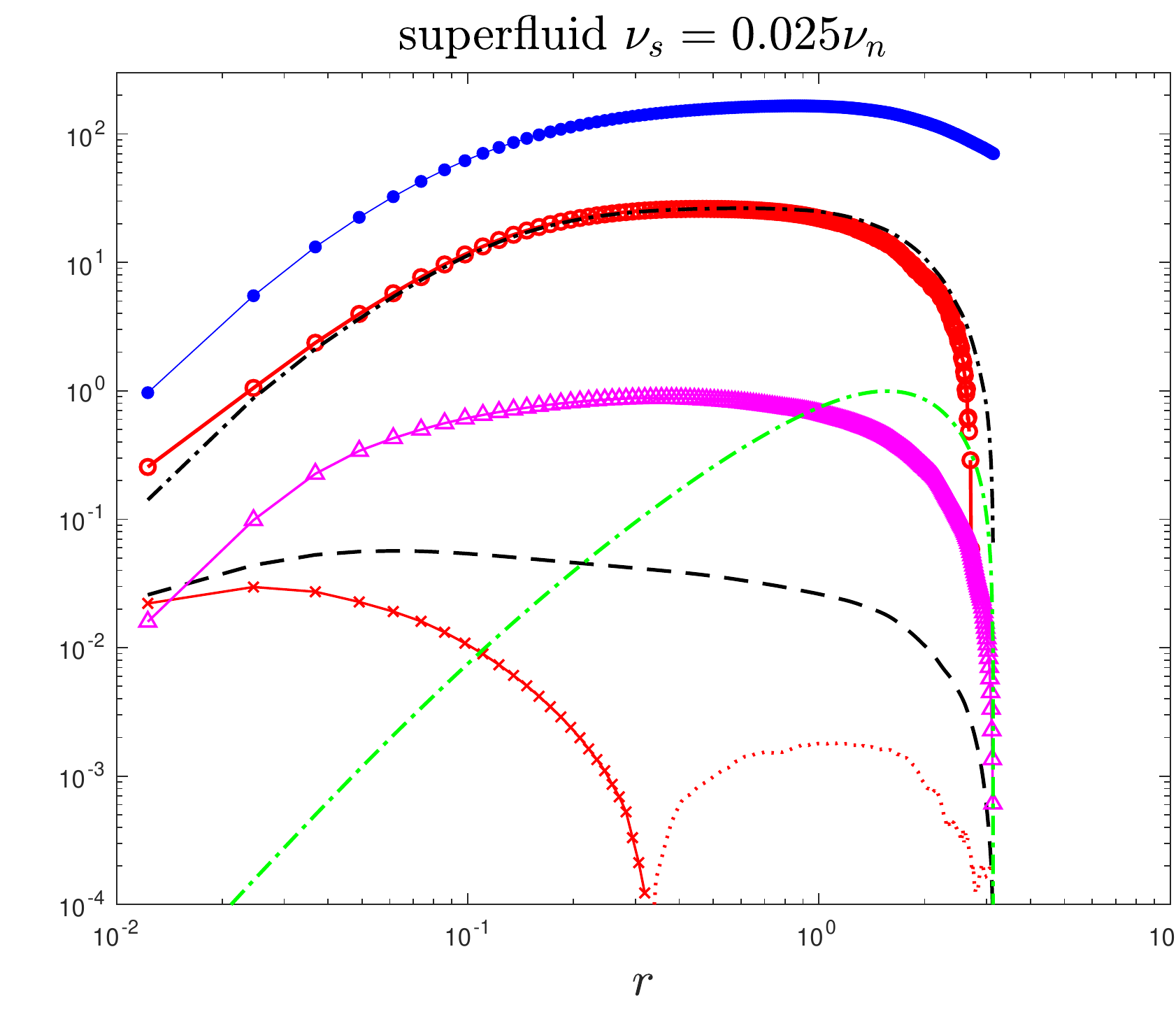}
\end{subfigure}%
\caption{Terms in the budget equations for the normal fluid  (left column) and for the superfluid  (right column).   Simulations are performed for density ratios $\rho_n/\rho = 0.5$ and for viscosity ratios:  $\nu_s/\nu_n=0.1$ for the top row and $\nu_s/\nu_n=0.025$ for the bottom row.
	All terms are normalized by $\varepsilon_*^{5/4}\nu_n^{1/4}$, with $\varepsilon_*=7e-4$ the constant energy rate injected to force turbulence for both fluid fractions.  We use the same legend as in corpus of the manuscript for  different terms in the equations.}
\label{fig:app_D3-balance}
\end{figure}

Figure \ref{fig:app_D3-balance} shows terms in the equations for different viscosity ratios. Noticeable  is the fact that variations of the density ratio affect only the dissipation source term of the superfluid, viz.  (- x) and (- -) on the right column. The inertial terms are the same. We can therefore  conclude that the choice of the viscosity ratio has limited influence on the results,  as long as the ratio of viscosities is small ($<0.1$) and respects the concept of the HVBK two-fluid model.  The same result is supported by our simulations for other values of density ratios. \\
For lowest temperatures, the resolution currently used is not sufficient to capture the smallest scales motion. We recall that the mesh size should be smaller than both the normal fluid Kolmogorov scale $\eta_n$, and the inter-vortex length of the superfluid. The superfluid also has its Kolmogorov scale $\eta_s$ but it should be irrelevant in the framework of  the HVBK two-fluid model.

\newpage

\section{Appendix 2. The effect of the resolution on the results} \label{appendix_resolution}

\begin{figure}[!h]
\begin{subfigure}{0.5\textwidth}
	\includegraphics[width=.75\textwidth,height=0.6\textheight,keepaspectratio,trim={1.5cm 7.2cm 2.4cm 7.5cm}, clip]{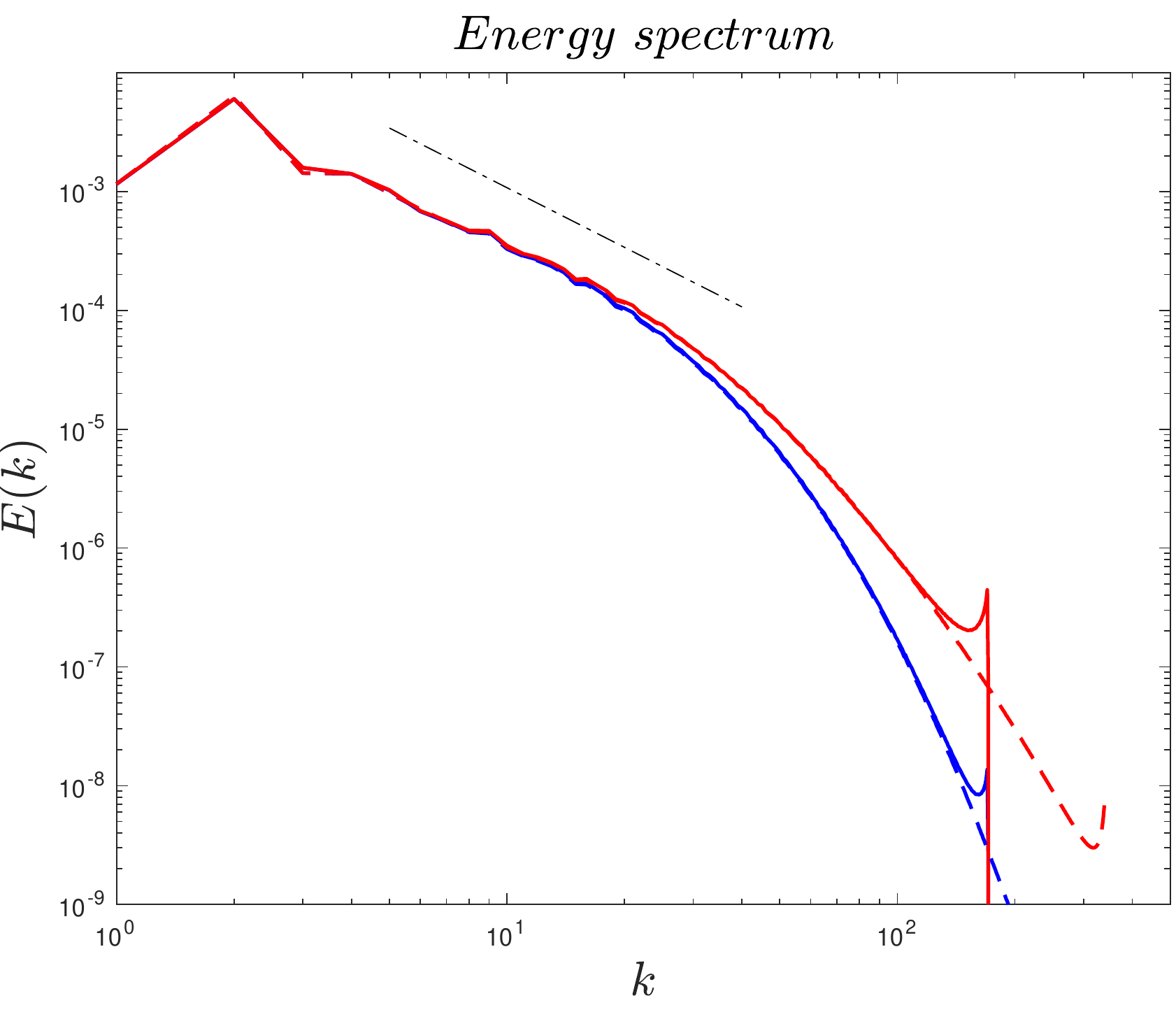}
\end{subfigure}
\begin{subfigure}{0.5\textwidth}
	\includegraphics[width=.75\textwidth,height=0.6\textheight,keepaspectratio,trim={1.5cm 7.2cm 2.4cm 7.5cm}, clip]{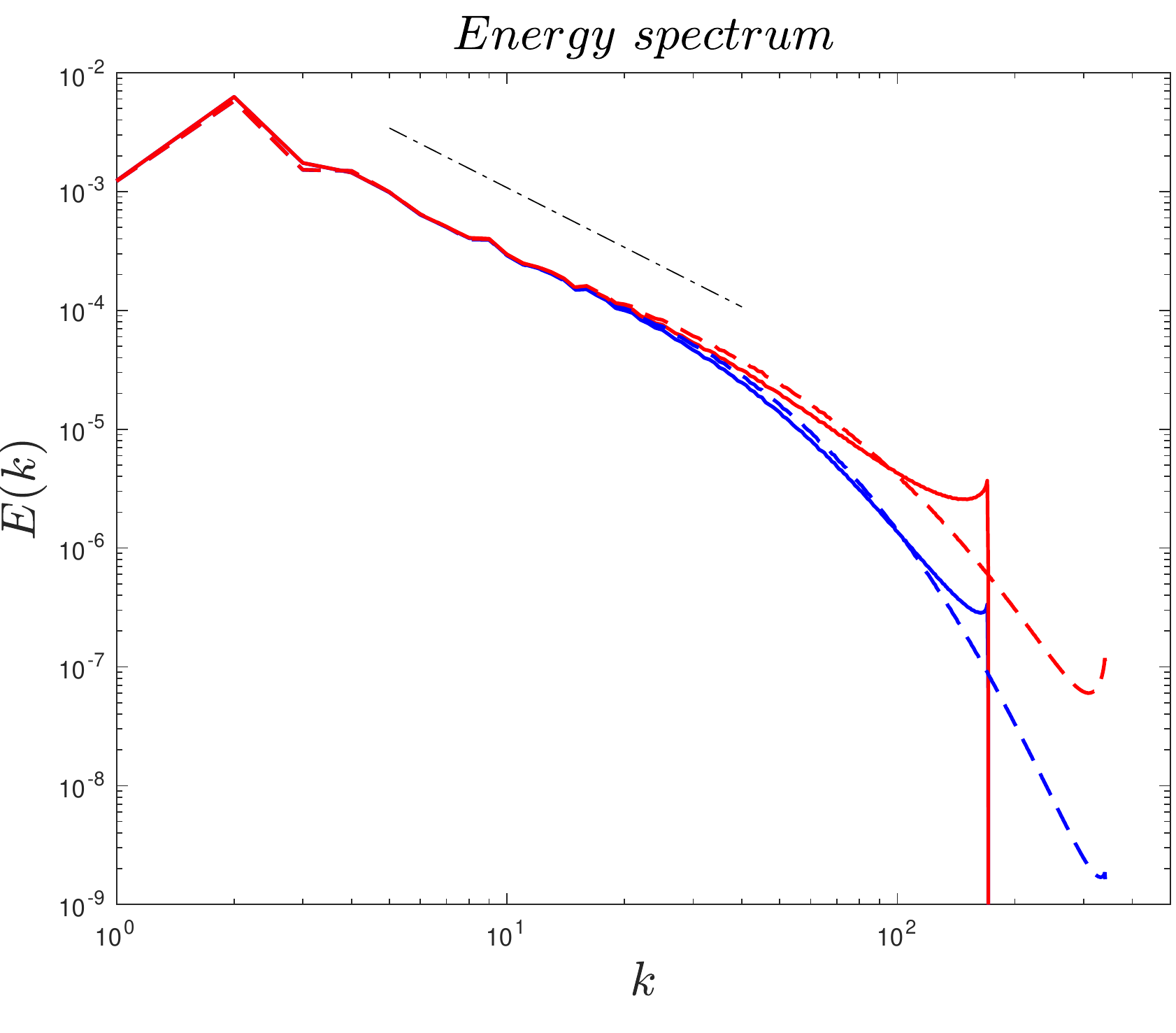}
\end{subfigure}%
\caption{The spectrum (left) for $\rho_s/\rho_n = 1$, (right) for $\rho_s/\rho_n = 10$. For different resolution (--) N=512, $k_{max}\eta_n = (N/3)\eta_n=1.816$, $k_{max}\eta_s = (N/3)\eta_s=0.4651$. (- -) N=1024, $k_{max}\eta_n = (N/3)\eta_n=3.8105$, $k_{max}\eta_s = (N/3)\eta_s=0.9132$.}
\label{fig:Ek-1024}
\end{figure}

\begin{figure}
\begin{subfigure}{0.5\textwidth}
	\includegraphics[width=.75\textwidth,height=0.6\textheight,keepaspectratio,trim={1.5cm 7.2cm 2.4cm 7.5cm}, clip]{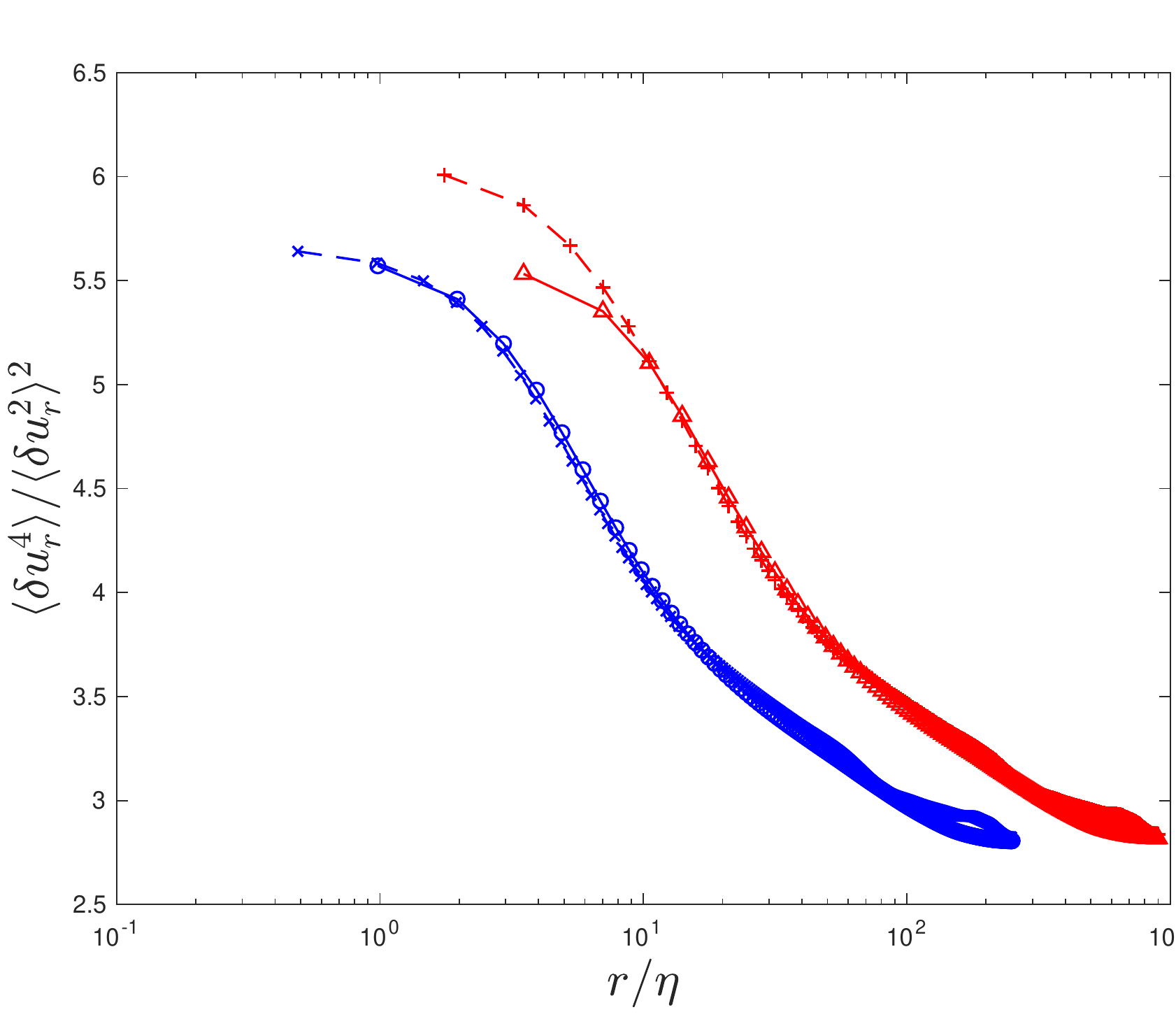}
\end{subfigure}%
\begin{subfigure}{0.5\textwidth}
	\includegraphics[width=.75\textwidth,height=0.6\textheight,keepaspectratio,trim={1.5cm 7.2cm 2.4cm 7.5cm}, clip]{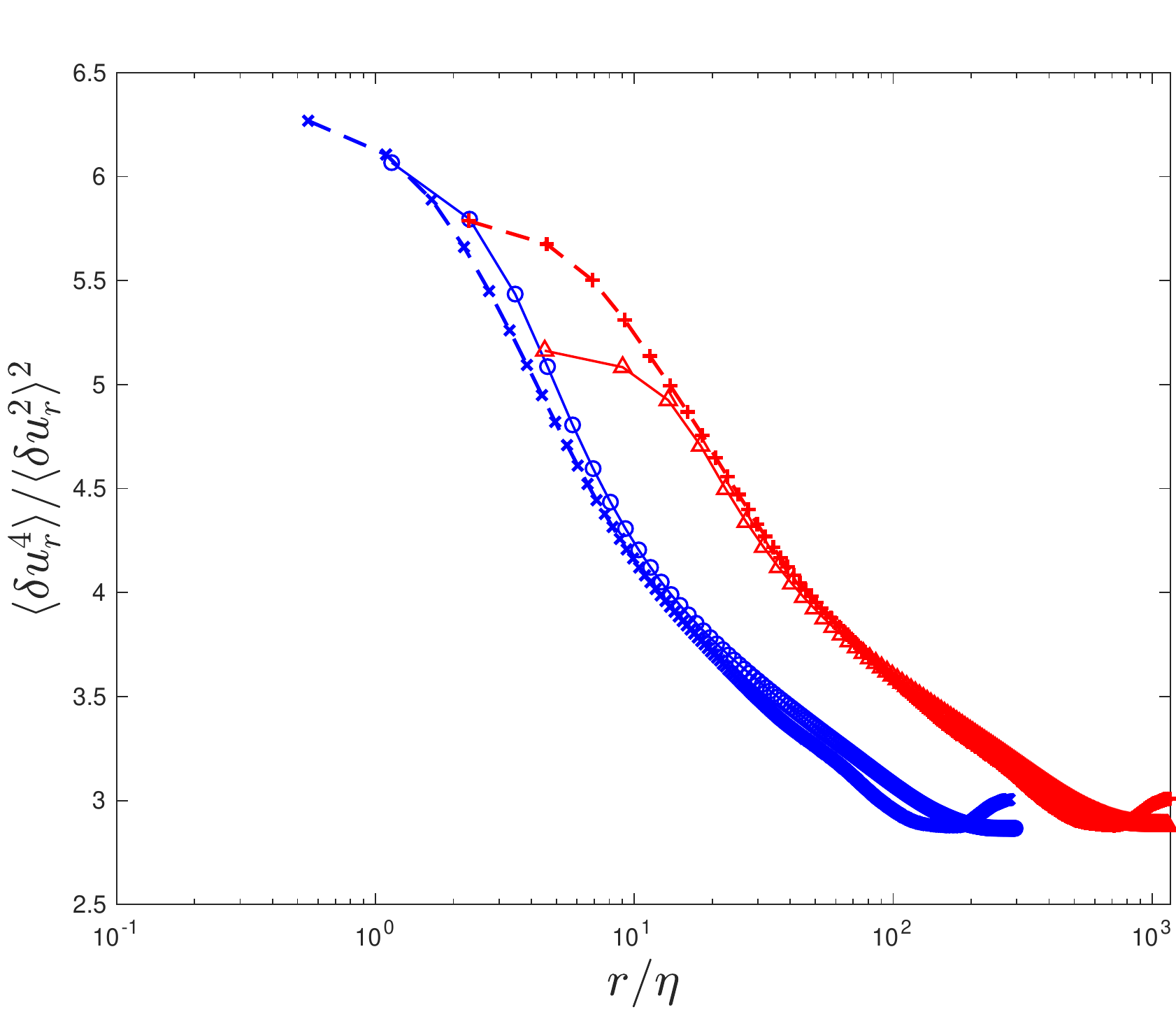}
\end{subfigure}%
\caption{The normalised "4th"-order structure function,  as a  function of $r/\eta_{n,s}$ for normal fluid in blue colour and superfluid in red colour. (left) for $\rho_s/\rho_n = 1$, (right) for $\rho_s/\rho_n = 10$. Resolutions (--) N=512, $k_{max}\eta_n = (N/3)\eta_n=1.816$, $k_{max}\eta_s = (N/3)\eta_s=0.4651$. (- -) N=1024, $k_{max}\eta_n = (N/3)\eta_n=3.8105$, $k_{max}\eta_s = (N/3)\eta_s=0.9132$.}
\label{fig:S4}
\end{figure}

We have performed additional numerical simulations, with a better resolution. Different statistics, such as  spectra  for  a number of grid points $N=512$ and $N=1024$ are depicted  in figure \ref{fig:Ek-1024} for $\rho_n/\rho_s = 1$ and $\rho_n/\rho_s = 0.1$ respectively. For  $N=1024$, the spectrum  is cut-off at higher wave numbers. The additional kinetic energy is, however, negligible. \\
The normalised 4th-order structure function (Fig. \ref{fig:S4}) tends towards a slightly  larger value at the smallest scales,  but still  within the error bars.  Figure \ref{fig:S4:derivative} depicts values for the flatness of the velocity derivative for two resolutions. While for the normal fluid the results are the same, we notice a slight increase of the flatness of the superfluid at the lowest temperature, from a value of $5.25$ obtained for $N=512$ to a value of $6$ for $N=1024$. 


\newpage

\section{Appendix 3.  Effect of considering the full expression of the mutual friction force}  \label{appendix_mutualfriction}

\begin{figure}[!h]
\begin{subfigure}{0.45\textwidth}
	\includegraphics[width=1.\textwidth,height=0.7\textheight,keepaspectratio,trim={1.5cm 7.2cm 2.4cm 7.5cm}, clip]{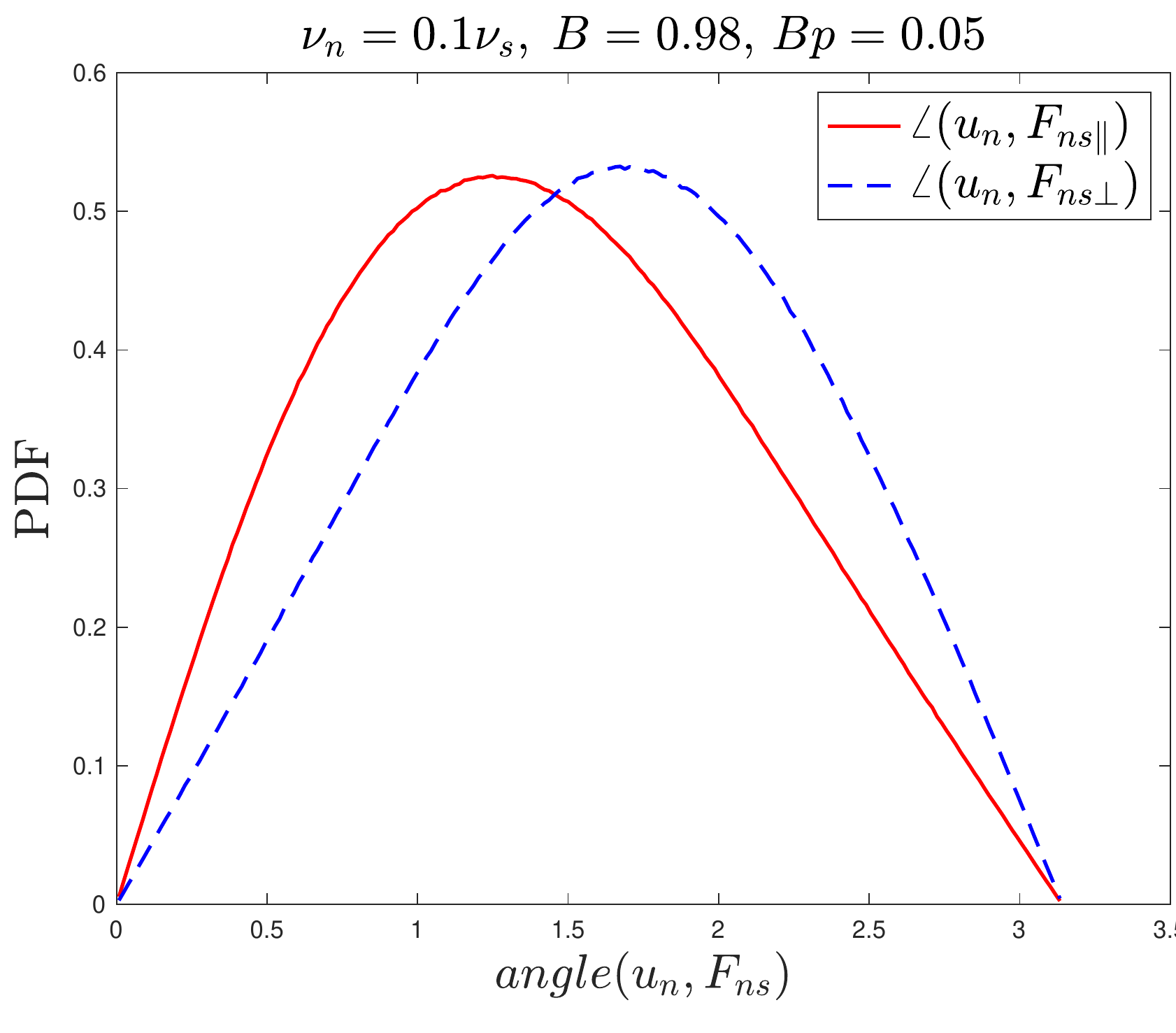}
\end{subfigure}%
\begin{subfigure}{0.45\textwidth}
	\includegraphics[width=1.\textwidth,height=0.7\textheight,keepaspectratio,trim={1.5cm 7.2cm 2.4cm 7.5cm}, clip]{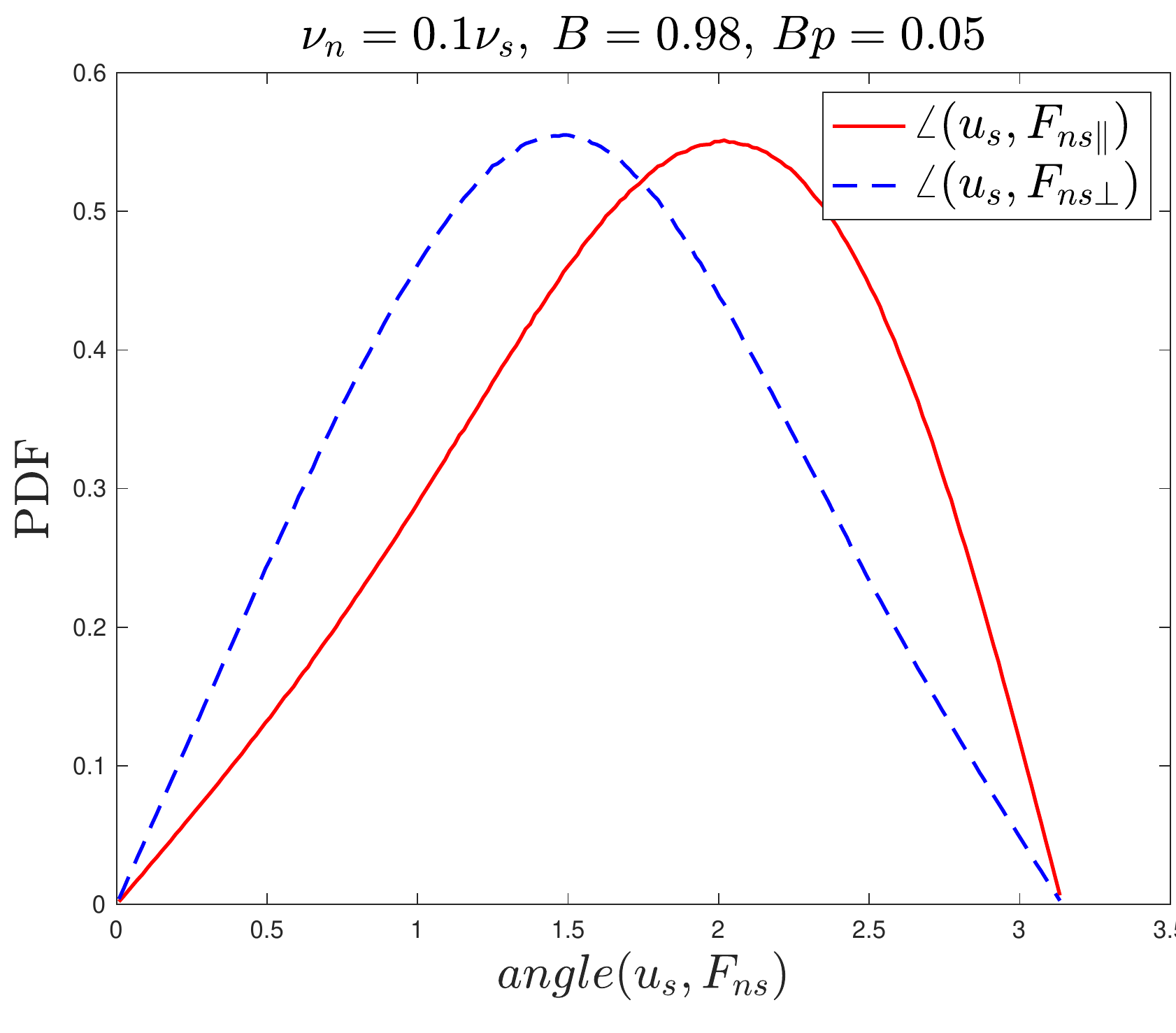}
\end{subfigure}%
\caption{The PDF of angle between the velocity and different friction force components.  Normal fluid  (left), and for the superfluid (right).}
\label{fig:pdf_angle_velocity_force}
\end{figure}

\begin{figure}[!h]
	\centering
\includegraphics[width=.5\textwidth,height=0.35\textheight,keepaspectratio,trim={1cm 7.2cm 2.4cm 7.5cm}, clip]{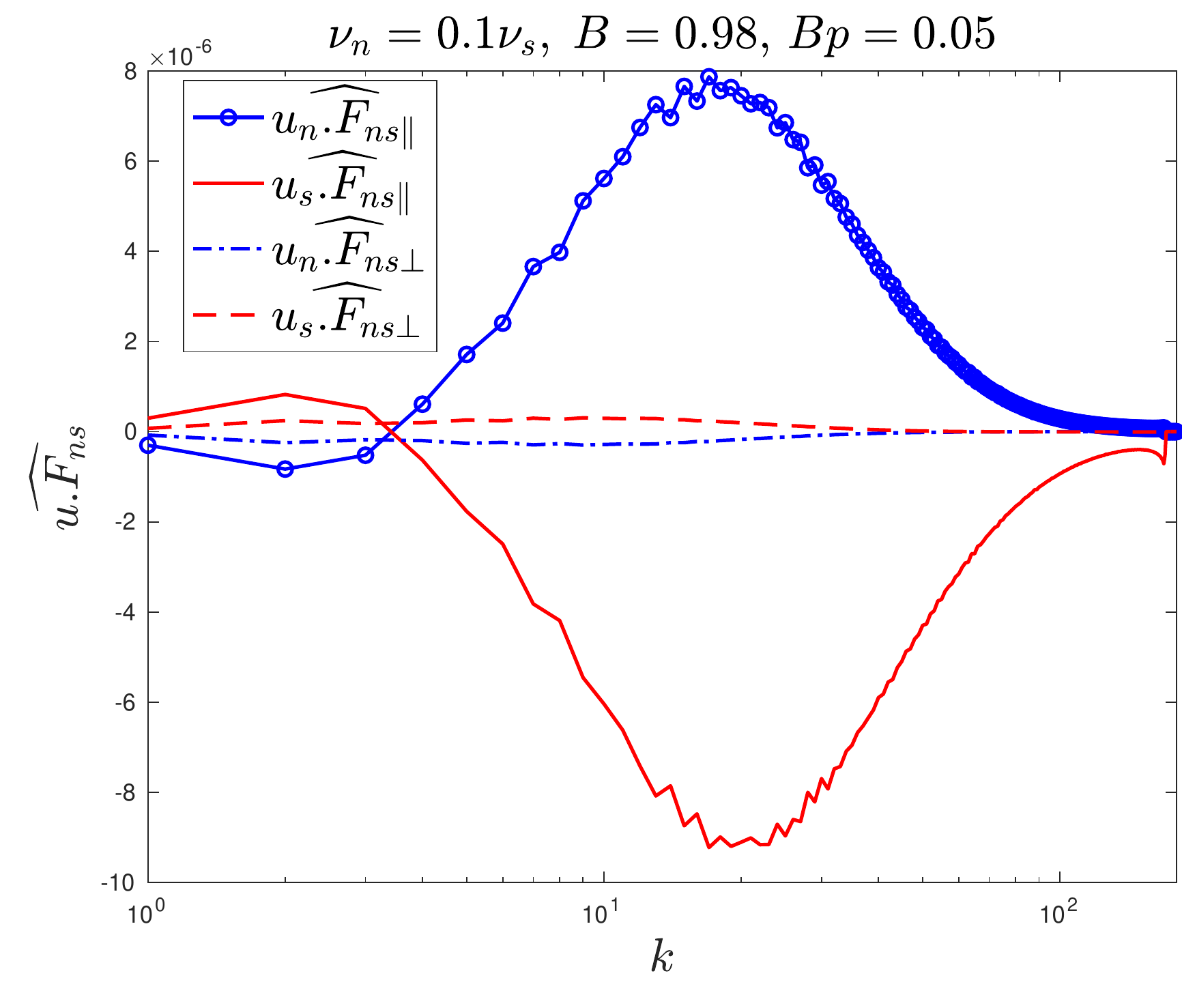}
\caption{The spectrum of (-o-) $\bm{u}_{n}F_{ns\parallel}$, (--) $\bm{u}_{s}F_{ns\parallel}$, (-.) $\bm{u}_{n}F_{ns\perp}$, (- -) $\bm{u}_{s}F_{ns\perp}$.}
\label{fig:spec_uFns}
\end{figure}

We test our claims by considering a generalized expression of the mutual friction force \citep{henderson2004superfluid}, viz. 
\begin{equation}
	\bm{F}_{ns} = \frac{B}{2}\frac{\rho_s\rho_n}{\rho}\widehat{\bm{\omega}}_s\times \left[ \bm{\omega_s}  \times  (\bm{v}_n -\bm{v}_s -\Dot{L})\right]+\frac{Bp}{2}\frac{\rho_s\rho_n}{\rho} \bm{\omega_s}  \times  (\bm{v}_n -\bm{v}_s - \Dot{L}),
	\label{eq:FM-full}
\end{equation}
where $\Dot{L}$ is the vortex line velocity due to the oscillation of the vortex wave and writes
\begin{equation}
	\Dot{L} = \frac{\kappa}{4\pi}log(l/a_0) \nabla \times \widehat{\bm{\omega}}_s
	\label{eq:dot_l}
\end{equation}
$l=\sqrt{\kappa/\bm{\omega}_s}$ is the inter-vortex length, $a_0$ is the vortex core size, and $\kappa$ is the unit circulation. \\ 
In this work, we have made two simplifications. 
First, we neglected the velocity due to the vortex line oscillation $\Dot{L}$. Because in the limit of  high Reynolds numbers, $l \rightarrow \mathcal{O}(a_0)$ implies $\Dot{L} \rightarrow \mathcal{O}(\kappa)$. The latter  is too small compared to the characteristic velocity of the superfluid to be taken into account. \\
On the other hand, in the original idea of the mutual friction force proposed by Hall and Vinen, the vortex lines are considered as filaments with no mass, which implies that the inertial effects of the vortex lines are irrelevant. As a matter of fact, based on the concept of two-fluid model, the superfluid velocity resolved by the NS equations is a space-smoothed value $\tilde{\bm{v}}_s$, which is the velocity induced by the vortex line smoothed (or averaged) over a large volume of fluid. In this context, $\tilde{\bm{v}}_s$ is equivalent to $\bm{v}_s + \Dot{L}$. The velocity due to the vortex tangle oscillation is not actually neglected, but merged into  $\tilde{\bm{v}}_s$. \\
Although we have ignored $\Dot{L}$ into a simplified scenario, we do not suggest $\Dot{L}$ should always be neglected. The contribution of $\Dot{L}$ is interesting to be considered in some situations. For instance, when  $\bm{v}_n -\bm{v}_s = 0$, due to the contribution of $\Dot{L}$ the mutual friction force is not zero.   However,  for the present work, we neglect $\Dot{L}$.
The mutual friction force then consists of two components: one is parallel to the relative velocity $\bm{v}_n -\bm{v}_s$, noted $F_{ns\parallel} = \frac{B}{2}\frac{\rho_s\rho_n}{\rho}\widehat{\bm{\omega}}_s\times \left[ \bm{\omega_s}  \times  (\bm{v}_n -\bm{v}_s)\right]$. The other one is perpendicular to the relative velocity $\bm{v}_n -\bm{v}_s$, noted $F_{ns\perp} = \frac{Bp}{2}\frac{\rho_s\rho_n}{\rho} \bm{\omega_s}  \times  (\bm{v}_n -\bm{v}_s)$. The mutual friction force is 
\begin{equation}
	\bm{F}_{ns} = \frac{B}{2}\frac{\rho_s\rho_n}{\rho}\widehat{\bm{\omega}}_s\times \left[ \bm{\omega_s}  \times  (\bm{v}_n -\bm{v}_s)\right]+\frac{Bp}{2}\frac{\rho_s\rho_n}{\rho} \bm{\omega_s}  \times  (\bm{v}_n -\bm{v}_s), 
	\label{eq:FM-full2}
\end{equation}
and with the supposition that  $\bm{\omega}_s \perp (\bm{v}_n -\bm{v}_s)$, it becomes 
\begin{equation}
	\bm{F}_{ns\parallel} = \frac{B}{2}\frac{\rho_s\rho_n}{\rho}\widehat{\bm{\omega}}_s\times \left[ \bm{\omega_s}  \times  (\bm{v}_n -\bm{v}_s)\right] = - \frac{B}{2}\frac{\rho_s\rho_n}{\rho}|\bm{\omega}_s|(\bm{v}_n -\bm{v}_s), 
	\label{eq:FM-parallel}
\end{equation}
and 
\begin{equation}
	\bm{F}_{ns\perp} = \frac{Bp}{2}\frac{\rho_s\rho_n}{\rho} \bm{\omega_s}  \times  (\bm{v}_n -\bm{v}_s).
	\label{eq:FM-perp}
\end{equation}
The second simplification of the present work is to consider only the component  $F_{ns\parallel}$,  basically because  $F_{ns\perp}$ is considered as being non-dissipative and represents a Magnus effect associated with quantized vortices  \citep{Roche_etal_2009}. It signifies that  $F_{ns\parallel}$ is responsible for  the energy exchange between the two components,  while $F_{ns\perp}$ does not contribute much to the energy exchange between the two fluids. \\
A first validation  of our considerations is backed by the statistics of the angle made by  $\bm{u}_n$ and $F_{ns}$. Figure \ref{fig:pdf_angle_velocity_force} depicts the PDF of the angle made by the velocity vector   and different components  of the force $F_{ns\perp}$ and $F_{ns\parallel}$, for the normal fluid (left) and the superfluid (right). The PDF $\measuredangle (\bm{u}_n, F_{ns\parallel})$ is skewed towards values of the angle between $(0,\pi/2)$. This signifies that $\bm{u}_n$ is preferably aligned with $F_{ns\parallel}$. Therefore,  $F_{ns\parallel}$ injects energy to the normal fluid. The PDF of the angle $\measuredangle (\bm{u}_n, F_{ns\perp})$ is  almost symmetric about the value of  $\pi/2$. This signifies that the $F_{ns\perp}$ does not inject energy to the normal fluid. 
The same qualitative observation holds for  the PDF of $\measuredangle (\bm{u}_s, F_{ns})$. The parallel component $F_{ns\parallel}$ extracts energy from the superfluid and $F_{ns\perp}$ does not affect, on average, the superfluid. These are arguments that serve as a basis in neglecting  $F_{ns\perp}$. \\
Furthermore, the temperature-related coefficient $B$ is generally larger than $Bp$. For example, for $T=1.95K$, $B=0.98$ and $Bp=0.05$. The spectrum of $\bm{u}_{n,s}F_{ns\perp}$ is negligible compared to that of $\bm{u}_{n,s}F_{ns\parallel}$, see figure \ref{fig:spec_uFns}. This is an additional reason for considering  the simplified form of the mutual friction force,  as provided by Eq. \eqref{eq:FM-parallel}, and considered  in the present work. \\ 
The scale-by-scale transport equation for the third--order structure function is not affected by the  consideration of the complete expression of the friction force,  \eqref{eq:FM-full2}. 
Additional numerical studies considering  the full expression of vortex oscillations will be performed in the future.

\newpage




\begin{thebibliography}{64}
	\expandafter\ifx\csname natexlab\endcsname\relax\def\natexlab#1{#1}\fi
	\def\au#1{#1} \def\ed#1{#1} \def\yr#1{#1}\def\at#1{#1}\def\jt#1{\textit{#1}}
	\def\bt#1{#1}\def\bvol#1{\textbf{#1}} \def\vol#1{#1} \def\pg#1{#1}
	\def\publ#1{#1}\def\arxiv#1{#1}\def\org#1{#1}\def\st#1{\textit{#1}}
	
	\bibitem[Abid {\em et~al.\/}(2003)Abid, Huepe, Metens, Nore, Pham, Tuckerman \&
	Brachet]{Abid2003509}
	{\sc \au{Abid, M.}, \au{Huepe, C.}, \au{Metens, S.}, \au{Nore, C.}, \au{Pham,
			C.}, \au{Tuckerman, L.} \& \au{Brachet, M.~E.}} \yr{2003}
	\at{{G}ross-{P}itaevskii dynamics of {B}ose-{E}instein condensates and
		superfluid turbulence}.  \jt{Fluid Dynamics Research}  \bvol{33}~(5-6),
	\pg{509--544}.
	
	\bibitem[Baggaley {\em et~al.\/}(2012)Baggaley, Barenghi, Shukurov \&
	Sergeev]{Baggaley_2012}
	{\sc \au{Baggaley, A.~W.}, \au{Barenghi, C.~F.}, \au{Shukurov, A.} \&
		\au{Sergeev, Y.~A.}} \yr{2012}  \at{Coherent vortex structures in quantum
		turbulence}.  \jt{Europhysics Lett.}  \bvol{98}~(2),  \pg{26002}.
	
	\bibitem[Balibar(2017)]{QT-Balibar-Tisza}
	{\sc \au{Balibar, S.}} \yr{2017}  \at{Laszlo {T}isza and the two-fluid model of
		superfluidity}.  \jt{Comptes Rendus Physique}  \bvol{18}~(9),  \pg{586--591}.
	
	\bibitem[Barenghi {\em et~al.\/}(1983)Barenghi, Donnelly \&
	Vinen]{Barenghi_1983}
	{\sc \au{Barenghi, C.~F.}, \au{Donnelly, R.~J.} \& \au{Vinen, W.~F.}} \yr{1983}
	\at{Friction on quantized vortices in {H}elium {II}. {A} review}.  \jt{J.
		Low Temp. Phys.}  \bvol{52},  \pg{189--247}.
	
	\bibitem[Barenghi {\em et~al.\/}(2014{\natexlab{{\em a\/}}})Barenghi,
	L{\textquoteright}vov \& Roche]{Barenghi_2013}
	{\sc \au{Barenghi, C.~F.}, \au{L{\textquoteright}vov, V.~S.} \& \au{Roche,
			P.-E.}} \yr{2014{\natexlab{{\em a\/}}}}  \at{Experimental, numerical, and
		analytical velocity spectra in turbulent quantum fluid}.  \jt{Proceedings of
		the National Academy of Sciences}  \bvol{111}~(Supplement 1),
	\pg{4683--4690}.
	
	\bibitem[Barenghi {\em et~al.\/}(2014{\natexlab{{\em b\/}}})Barenghi, Skrbek \&
	Sreenivasan]{Barenghi_etal_2014}
	{\sc \au{Barenghi, C.~F.}, \au{Skrbek, L.} \& \au{Sreenivasan, K.~R.}}
	\yr{2014{\natexlab{{\em b\/}}}}  \at{Introduction to quantum turbulence}.
	\jt{Proc. Natl. Acad. Sci.}  \bvol{111},  \pg{4647}.
	
	\bibitem[Batchelor \& Townsend(1949)]{Batchelor_Townsend_nature_1949}
	{\sc \au{Batchelor, G.~K.} \& \au{Townsend, A.~A.}} \yr{1949}  \at{The nature
		of turbulent motion at large wave-numbers}.  \jt{Proceedings of the Royal
		Society of London. Series A. Mathematical and Physical Sciences}  \bvol{199},
	\pg{238--255}.
	
	\bibitem[Biferale {\em et~al.\/}(2018)Biferale, Khomenko, L'vov, Pomyalov,
	Procaccia \& Sahoo]{Biferale_2018}
	{\sc \au{Biferale, L.}, \au{Khomenko, D.}, \au{L'vov, V.~S.}, \au{Pomyalov,
			A.}, \au{Procaccia, I.} \& \au{Sahoo, G.}} \yr{2018}  \at{Turbulent
		statistics and intermittency enhancement in coflowing superfluid
		$^{4}\mathrm{He}$}.  \jt{Phys. Rev. Fluids}  \bvol{3},  \pg{024605}.
	
	\bibitem[Boschung {\em et~al.\/}(2017)Boschung, Hennig, Denker, Pitsch \&
	Hill]{Boschung2017}
	{\sc \au{Boschung, J.}, \au{Hennig, F.}, \au{Denker, D.}, \au{Pitsch, H.} \&
		\au{Hill, R.~J.}} \yr{2017}  \at{Analysis of structure function equations up
		to the seventh order}.  \jt{J. of Turbulence}  \bvol{18}~(11),
	\pg{1001--1032}.
	
	\bibitem[Bou\'e {\em et~al.\/}(2015)Bou\'e, L'vov, Nagar, Nazarenko, Pomyalov
	\& Procaccia]{Boue_2015}
	{\sc \au{Bou\'e, L.}, \au{L'vov, V.~S.}, \au{Nagar, Y.}, \au{Nazarenko, S.~V.},
		\au{Pomyalov, A.} \& \au{Procaccia, I.}} \yr{2015}  \at{Energy and vorticity
		spectra in turbulent superfluid $^{4}\mathrm{He}$ from $\mathrm{T}=0$ to
		${T}_{\ensuremath{\lambda}}$}.  \jt{Phys. Rev. B}  \bvol{91},  \pg{144501}.
	
	\bibitem[Bou\'e {\em et~al.\/}(2013)Bou\'e, L'vov, Pomyalov \&
	Procaccia]{Boue_2013}
	{\sc \au{Bou\'e, L.}, \au{L'vov, V.~S.}, \au{Pomyalov, A.} \& \au{Procaccia,
			I.}} \yr{2013}  \at{Enhancement of intermittency in superfluid turbulence}.
	\jt{Phys. Rev. Lett.}  \bvol{110},  \pg{014502}.
	
	\bibitem[Bradley {\em et~al.\/}(2008)Bradley, Fisher, Gu\'enault, Haley,
	O'Sullivan, Pickett \& Tsepelin]{Bradley_2008}
	{\sc \au{Bradley, D.~I.}, \au{Fisher, S.~N.}, \au{Gu\'enault, A.~M.},
		\au{Haley, R.~P.}, \au{O'Sullivan, S.}, \au{Pickett, G.~R.} \& \au{Tsepelin,
			V.}} \yr{2008}  \at{Fluctuations and correlations of pure quantum turbulence
		in superfluid $^{3}\mathrm{He}\mathrm{\text{\ensuremath{-}}}\mathrm{B}$}.
	\jt{Phys. Rev. Lett.}  \bvol{101}~(4),  \pg{065302}.
	
	\bibitem[Djenidi {\em et~al.\/}(2017{\natexlab{{\em a\/}}})Djenidi, Antonia \&
	Danaila]{Djenidi2017PRF}
	{\sc \au{Djenidi, L.}, \au{Antonia, R.~A.} \& \au{Danaila, L.}}
	\yr{2017{\natexlab{{\em a\/}}}}  \at{Self-preservation relation to the
		\uppercase{K}olmogorov similarity hypotheses}.  \jt{Phys. Rev. Fluids}
	\bvol{2},  \pg{054606}.
	
	\bibitem[Djenidi {\em et~al.\/}(2017{\natexlab{{\em b\/}}})Djenidi, Antonia,
	Danaila \& Tang]{Lyazid2017}
	{\sc \au{Djenidi, L.}, \au{Antonia, R.~A.}, \au{Danaila, L.} \& \au{Tang,
			S.~L.}} \yr{2017{\natexlab{{\em b\/}}}}  \at{A note on the velocity
		derivative flatness factor in decaying \uppercase{HIT}}.  \jt{Phys. of
		Fluids}  \bvol{29},  \pg{051702}.
	
	\bibitem[Donnelly(1991)]{QT-book-1991-donnelly}
	{\sc \au{Donnelly, R.~J.}}, ed. \yr{1991} {\em Quantized Vortices in Helium
		II\/}.  \publ{CUP}.
	
	\bibitem[Donnelly(2009)]{2009_Donnelly_yearophysic}
	{\sc \au{Donnelly, R.~J.}} \yr{2009}  \at{The two-fluid theory and second sound
		in liquid \uppercase{H}elium}.  \jt{Physics Today}  \bvol{62}~(10),
	\pg{34--39}.
	
	\bibitem[Galantucci {\em et~al.\/}(2020)Galantucci, Baggaley, Barenghi \&
	Krstulovic]{QT-coupling-2020-gal}
	{\sc \au{Galantucci, L.}, \au{Baggaley, A.~W.}, \au{Barenghi, C.~F.} \&
		\au{Krstulovic, G.}} \yr{2020}  \at{A new self-consistent approach of quantum
		turbulence in superfluid helium}.  \jt{The European Physical Journal Plus}
	\bvol{135}~(7),  \pg{547}.
	
	\bibitem[Galantucci {\em et~al.\/}(2021)Galantucci, Krstulovic \&
	Barenghi]{galantucci2021-bundle}
	{\sc \au{Galantucci, L.}, \au{Krstulovic, G.} \& \au{Barenghi, C.~F.}}
	\yr{2021}  \at{Friction-enhanced lifetime of bundled quantum vortices}.
	\jt{arXiv}  \bvol{2107.07768}.
	
	\bibitem[Gauding {\em et~al.\/}(2017)Gauding, Danaila \&
	Varea]{gauding2017high}
	{\sc \au{Gauding, M.}, \au{Danaila, L.} \& \au{Varea, E.}} \yr{2017}
	\at{High-order structure functions for passive scalar fed by a mean
		gradient}.  \jt{International Journal of Heat and Fluid Flow}  \bvol{67},
	\pg{86--93}.
	
	\bibitem[Gotoh \& Nakano(2003)]{Gotoh_Nakano_2003}
	{\sc \au{Gotoh, T.} \& \au{Nakano, T.}} \yr{2003}  \at{Role of pressure in
		turbulence}.  \jt{J. Stat. Phys.}  \bvol{113},  \pg{855--875}.
	
	\bibitem[Hall \& Vinen(1956)]{Hall_Vinen_1956}
	{\sc \au{Hall, H.~E.} \& \au{Vinen, W.~F.}} \yr{1956}  \at{The rotation of
		liquid $\rm{Helium}$ $\mathrm{II}$: Experiments on the propagation of second
		sound in uniformly rotating $\rm{Helium}$ $\mathrm{II}$}.  \jt{Proc. Roy.
		Soc. London}  \bvol{Ser A, 238},  \pg{204}.
	
	\bibitem[Halperin \& Tsubota(2009)]{QT-book-2009-tsubota}
	{\sc \au{Halperin, B.} \& \au{Tsubota, M.}}, ed. \yr{2009} {\em Quantum
		Turbulence\/}. {\em Progress in Low Temperature Physics\/} 16.
	\publ{Springer}.
	
	\bibitem[Henderson \& Barenghi(2004)]{henderson2004superfluid}
	{\sc \au{Henderson, K.~L.} \& \au{Barenghi, C.~F.}} \yr{2004}  \at{Superfluid
		couette flow in an enclosed annulus}.  \jt{Theoretical and Computational
		Fluid Dynamics}  \bvol{18}~(2),  \pg{183--196}.
	
	\bibitem[Hill(2001)]{Hill2001}
	{\sc \au{Hill, R.}} \yr{2001}  \at{Equations relating structure functions of
		all orders.}  \jt{J. Fluid Mech.}  \bvol{434},  \pg{379--388}.
	
	\bibitem[Hill \& Boratav(2001)]{Hill_Boratav2001}
	{\sc \au{Hill, R.} \& \au{Boratav, O.}} \yr{2001}  \at{Next-order structure
		function equations}.  \jt{Phys. of Fluids}  \bvol{13},  \pg{276}.
	
	\bibitem[Ishihara \& Gotoh(2009)]{Ishihara_2009review}
	{\sc \au{Ishihara, T.} \& \au{Gotoh, T.and~Kaneda, Y.}} \yr{2009}  \at{Study of
		high-{R}eynolds number isotropic turbulence by direct numerical simulation}.
	\jt{Annual Review of Fluid Mechanics}  \bvol{41}~(1),  \pg{165--180}.
	
	\bibitem[Ishihara {\em et~al.\/}(2007)Ishihara, Kaneda, Yokokawa, Itakura \&
	Uno]{Ishihara_2007}
	{\sc \au{Ishihara, T.}, \au{Kaneda, Y.}, \au{Yokokawa, M.}, \au{Itakura, K.} \&
		\au{Uno, A.}} \yr{2007}  \at{Small-scale statistics in high-resolution direct
		numerical simulation of turbulence: {R}eynolds number dependence of one-point
		velocity gradient statistics}.  \jt{J. Fluid Mech.}  \bvol{592},
	\pg{335--366}.
	
	\bibitem[Jou {\em et~al.\/}(2011)Jou, Mongiov{\`\i} \&
	Sciacca]{jou2011hydrodynamic}
	{\sc \au{Jou, D.}, \au{Mongiov{\`\i}, M.~S.} \& \au{Sciacca, M.}} \yr{2011}
	\at{Hydrodynamic equations of anisotropic, polarized and inhomogeneous
		superfluid vortex tangles}.  \jt{Physica D: Nonlinear Phenomena}
	\bvol{240}~(3),  \pg{249--258}.
	
	\bibitem[Khalatnikov(1965)]{QT-book-1965-khal}
	{\sc \au{Khalatnikov, I.~M.}} \yr{1965} {\em An Introduction to the Theory of
		Superfluidity\/}.  \publ{Benjamin}.
	
	\bibitem[Kobayashi {\em et~al.\/}(2021)Kobayashi, Parnaudeau, Luddens,
	Lothod{\'e}, Danaila, Brachet \& Danaila]{dan-2021-CPC-QUTE}
	{\sc \au{Kobayashi, M.}, \au{Parnaudeau, P.}, \au{Luddens, F.},
		\au{Lothod{\'e}, C.}, \au{Danaila, L.}, \au{Brachet, M.} \& \au{Danaila, I.}}
	\yr{2021}  \at{Quantum turbulence simulations using the {G}ross-{P}itaevskii
		equation: {H}igh-performance computing and new numerical benchmarks}.
	\jt{Computer Physics Communications}  \bvol{258},  \pg{107579}.
	
	\bibitem[Kolmogorov(1941{\natexlab{{\em a\/}}})]{Kolmogorov1941}
	{\sc \au{Kolmogorov, A.~N.}} \yr{1941{\natexlab{{\em a\/}}}}  \at{Dissipation
		of energy in the locally isotropic turbulence}.  \jt{Dokl. Akad. Nauk SSSR}
	\bvol{32(1)},  \pg{16--18}.
	
	\bibitem[Kolmogorov(1941{\natexlab{{\em b\/}}})]{Kolmogorov1941a}
	{\sc \au{Kolmogorov, A.~N.}} \yr{1941{\natexlab{{\em b\/}}}}  \at{The local
		structure of turbulence in incompressible viscous fluids for very large
		{R}eynolds numbers}.  \jt{Dokl. Akad. Nauk SSSR}  \bvol{30(4)},
	\pg{301--305}.
	
	\bibitem[Kolmogorov(1962)]{Kolmogorov1962}
	{\sc \au{Kolmogorov, A.~N.}} \yr{1962}  \at{A refinement of previous hypotheses
		concerning the local structure of turbulence in a viscous incompressible
		fluid at high {R}eynolds number}.  \jt{J. Fluid Mech.}  \bvol{13(1)},
	\pg{82--85}.
	
	\bibitem[Krstulovic(2016)]{krstulovic2016}
	{\sc \au{Krstulovic, G.}} \yr{2016}  \at{Grid superfluid turbulence and
		intermittency at very low temperature}.  \jt{Phys. Rev. E}  \bvol{93},
	\pg{063104}.
	
	\bibitem[Landau(1941)]{Landau1941}
	{\sc \au{Landau, L.}} \yr{1941}  \at{Theory of the superfluidity of
		\uppercase{H}elium \uppercase{II}}.  \jt{Physical Review}  \bvol{60}~(4),
	\pg{356--358}.
	
	\bibitem[Lipniacki(2006)]{QT-Lipniacki-2006}
	{\sc \au{Lipniacki, T.}} \yr{2006}  \at{Dynamics of superfluid 4{H}e: Two-scale
		approach}.  \jt{European Journal of Mechanics - B/Fluids}  \bvol{25}~(4),
	\pg{435--458}.
	
	\bibitem[Lvov {\em et~al.\/}(2006)Lvov, Nazarenko \& Skrbek]{Lvov_2006}
	{\sc \au{Lvov, V.}, \au{Nazarenko, S.} \& \au{Skrbek, L.}} \yr{2006}
	\at{Energy spectra of developed turbulence in $\mathrm{Helium}$ superfluids}.
	\jt{J. of Low Temperature Physics}  \bvol{145},  \pg{125--142}.
	
	\bibitem[Maurer \& Tabeling(1998)]{Maurer_1998}
	{\sc \au{Maurer, J.} \& \au{Tabeling, P.}} \yr{1998}  \at{Local investigation
		of superfluid turbulence}.  \jt{Europhysics Lett.}  \bvol{43}~(1),
	\pg{29--34}.
	
	\bibitem[Mongiovi {\em et~al.\/}(2018)Mongiovi, Jou \& Sciacca]{MONGIOVI2018}
	{\sc \au{Mongiovi, M.~S.}, \au{Jou, D.} \& \au{Sciacca, M.}} \yr{2018}
	\at{Non-equilibrium thermodynamics, heat transport and thermal waves in
		laminar and turbulent superfluid helium}.  \jt{Physics Reports}  \bvol{726},
	\pg{1--71}, non-equilibrium thermodynamics, heat transport and thermal waves
	in laminar and turbulent superfluid helium.
	
	\bibitem[Nemirovskii(2013)]{QT-review-2013-nemirovskii}
	{\sc \au{Nemirovskii, S.~K.}} \yr{2013}  \at{Quantum turbulence: Theoretical
		and numerical problems}.  \jt{Physics Reports}  \bvol{524},  \pg{85--202}.
	
	\bibitem[Nemirovskii(2020)]{nemirovskii2020}
	{\sc \au{Nemirovskii, S.~K.}} \yr{2020}  \at{On the closure problem of the
		coarse-grained hydrodynamics of turbulent superfluids}.  \jt{Journal of Low
		Temperature Physics}  \bvol{201},  \pg{254--268}.
	
	\bibitem[Nore {\em et~al.\/}(1997)Nore, Abid \& Brachet]{Nore97a}
	{\sc \au{Nore, C.}, \au{Abid, M.} \& \au{Brachet, M.~E.}} \yr{1997}
	\at{Decaying {K}olmogorov turbulence in a model of superflow}.  \jt{Physics
		of Fluids}  \bvol{9}~(9),  \pg{2644--2669}.
	
	\bibitem[Roberts \& Donnelly(1974)]{Roberts_1974}
	{\sc \au{Roberts, P.~H.} \& \au{Donnelly, R.~J.}} \yr{1974}  \at{Superfluid
		mechanics}.  \jt{Annual Review of Fluid Mechanics}  \bvol{6}~(1),
	\pg{179--225}.
	
	\bibitem[Roche {\em et~al.\/}(2009)Roche, Barenghi \&
	L\'ev\^eque]{Roche_etal_2009}
	{\sc \au{Roche, P.-E.}, \au{Barenghi, C.~F.} \& \au{L\'ev\^eque, E.}} \yr{2009}
	\at{Quantum turbulence at finite temperature: the two-fluids cascade}.
	\jt{European Phys. Lett.}  \bvol{87},  \pg{54006}.
	
	\bibitem[Roche {\em et~al.\/}(2007)Roche, Diribarne, Didelot, Fran{\c{c}}ais,
	Rousseau \& Willaime]{Roche_2007}
	{\sc \au{Roche, P.-E.}, \au{Diribarne, P.}, \au{Didelot, T.},
		\au{Fran{\c{c}}ais, O.}, \au{Rousseau, L.} \& \au{Willaime, H.}} \yr{2007}
	\at{Vortex density spectrum of quantum turbulence}.  \jt{Europhysics Lett.}
	\bvol{77}~(6),  \pg{66002}.
	
	\bibitem[Rusaouen {\em et~al.\/}(2017)Rusaouen, Chabaud, Salort \&
	Roche]{Rusaouen_etal_2017}
	{\sc \au{Rusaouen, E.}, \au{Chabaud, B.}, \au{Salort, J.} \& \au{Roche, P.-E.}}
	\yr{2017}  \at{Intermittency of quantum turbulence with superfluid fractions
		from $0\%$ to $96\%$}.  \jt{Phys. of Fluids}  \bvol{29},  \pg{105108}.
	
	\bibitem[Salort {\em et~al.\/}(2010{\natexlab{{\em a\/}}})Salort, Baudet,
	Castaing, Chabaud \& Daviaud]{Salort_etal_2010}
	{\sc \au{Salort, J.}, \au{Baudet, C.}, \au{Castaing, B.}, \au{Chabaud, B.} \&
		\au{Daviaud, F.}} \yr{2010{\natexlab{{\em a\/}}}}  \at{The rotation of liquid
		{H}elium {II}: Experiments on the propagation of second sound in uniformly
		rotating {H}elium {II}}.  \jt{Phys. of Fluids}  \bvol{22},  \pg{125102}.
	
	\bibitem[Salort {\em et~al.\/}(2010{\natexlab{{\em b\/}}})Salort, Baudet,
	Castaing, Chabaud, Daviaud, Didelot, Diribarne, Dubrulle, Gagne, Gauthier,
	Girard, H\'ebral, Rousset, Thibault \& Roche]{Salort2010Specmeasure}
	{\sc \au{Salort, J.}, \au{Baudet, C.}, \au{Castaing, B.}, \au{Chabaud, B.},
		\au{Daviaud, F.}, \au{Didelot, T.}, \au{Diribarne, P.}, \au{Dubrulle, B.},
		\au{Gagne, Y.}, \au{Gauthier, F.}, \au{Girard, A.}, \au{H\'ebral, B.},
		\au{Rousset, B.}, \au{Thibault, P.} \& \au{Roche, P.-E.}}
	\yr{2010{\natexlab{{\em b\/}}}}  \at{Turbulent velocity spectra in superfluid
		flows}.  \jt{Phys. of Fluids}  \bvol{22}~(12),  \pg{125102}.
	
	\bibitem[Salort {\em et~al.\/}(2012)Salort, Chabaud, L\'ev\^eque \&
	Roche]{Salort_etal_2012}
	{\sc \au{Salort, J.}, \au{Chabaud, B.}, \au{L\'ev\^eque, E.} \& \au{Roche,
			P.-E.}} \yr{2012}  \at{Energy cascade and the four-fifths law in superfluid
		turbulence}.  \jt{European Phys. Lett.}  \bvol{97},  \pg{34006}.
	
	\bibitem[Sasa {\em et~al.\/}(2011)Sasa, Kano, Machida, Lvov, Rudenko \&
	Tsubota]{Narimsa_2011}
	{\sc \au{Sasa, N.}, \au{Kano, T.}, \au{Machida, M.}, \au{Lvov, V.},
		\au{Rudenko, O.} \& \au{Tsubota, M.}} \yr{2011}  \at{Energy spectra of
		quantum turbulence: Large-scale simulation and modeling}.  \jt{Phys. Rev. B}
	\bvol{84},  \pg{054525}.
	
	\bibitem[She \& L\'ev\^eque(1994)]{She_1994universal}
	{\sc \au{She, Z.-S.} \& \au{L\'ev\^eque, E.}} \yr{1994}  \at{Universal scaling
		laws in fully developed turbulence}.  \jt{Phys. Rev. Lett.}  \bvol{72}~(3),
	\pg{336}.
	
	\bibitem[Shi(2021)]{Shi_qian_2021}
	{\sc \au{Shi, J.}} \yr{2021}  \at{Qian {Jian} (1939--2018) and his contribution
		to small-scale turbulence studies}.  \jt{Phys. of Fluids}  \bvol{33}~(4),
	\pg{041301}.
	
	\bibitem[Shukla \& Pandit(2016)]{Shukla2016}
	{\sc \au{Shukla, V.} \& \au{Pandit, R.}} \yr{2016}  \at{Multiscaling in
		superfluid turbulence: A shell-model study}.  \jt{Phys. Rev. E}  \bvol{94},
	\pg{043101}.
	
	\bibitem[Skrbek \& Sreenivasan(2012{\natexlab{{\em
				a\/}}})]{QT-review-sreeni-2012}
	{\sc \au{Skrbek, L.} \& \au{Sreenivasan, K.~R.}} \yr{2012{\natexlab{{\em
					a\/}}}}  \at{Developed quantum turbulence}.  \jt{Phys. Fluids}  \bvol{24},
	\pg{011301(1--48)}.
	
	\bibitem[Skrbek \& Sreenivasan(2012{\natexlab{{\em
				b\/}}})]{QT-book-sreeni-2012}
	{\sc \au{Skrbek, L.} \& \au{Sreenivasan, K.~R.}} \yr{2012{\natexlab{{\em
					b\/}}}}  \at{How similar is quantum turbulence to classical turbulence?}
	\bt{In {\em Ten Chapters in Turbulence\/} (ed. \ed{P.~A. Davidson, Y.~Kaneda
			\& K.~R. Sreenivasan})},  \pg{p. 405}.  \publ{Cambridge University Press}.
	
	\bibitem[Tang {\em et~al.\/}(2017)Tang, Antonia, Djenidi, Danaila \&
	Zhou]{Shunlin2017}
	{\sc \au{Tang, S.~L.}, \au{Antonia, R.~A.}, \au{Djenidi, L.}, \au{Danaila, L.}
		\& \au{Zhou, Y.}} \yr{2017}  \at{Finite \uppercase{R}eynolds number effect on
		the scaling range behavior of turbulent longitudinal velocity structure
		functions}.  \jt{J.~Fluid Mech.}  \bvol{820},  \pg{341--369}.
	
	\bibitem[Tang {\em et~al.\/}(2018)Tang, Antonia, Djenidi, Danaila \&
	Zhou]{Shunlin2018}
	{\sc \au{Tang, S.~L.}, \au{Antonia, R.~A.}, \au{Djenidi, L.}, \au{Danaila, L.}
		\& \au{Zhou, Y.}} \yr{2018}  \at{Reappraisal of the velocity derivative
		flatness factor in various turbulent flows}.  \jt{J.~Fluid Mech.}
	\bvol{847},  \pg{244--265}.
	
	\bibitem[Tisza(1938)]{Tisza_1938}
	{\sc \au{Tisza, L.}} \yr{1938}  \at{Transport phenomena in \uppercase{H}elium
		\uppercase{II}}.  \jt{Nature}  \bvol{141},  \pg{913}.
	
	\bibitem[Townsend(1951)]{Townsend1951fine}
	{\sc \au{Townsend, A.~A.}} \yr{1951}  \at{On the fine-scale structure of
		turbulence}.  \jt{Proc. R. Soc. Lond. A}  \bvol{208}~(1095),  \pg{534--542}.
	
	\bibitem[Tsubota {\em et~al.\/}(2017)Tsubota, Fujimoto \&
	Yui]{QT-review-2017-tsubota-num}
	{\sc \au{Tsubota, M.}, \au{Fujimoto, K.} \& \au{Yui, S.}} \yr{2017}
	\at{Numerical studies of quantum turbulence}.  \jt{J. of Low Temperature
		Physics}  \bvol{188},  \pg{119--189}.
	
	\bibitem[Vinen \& Niemela(2002)]{QT-review-2002-vinen}
	{\sc \au{Vinen, W.~F.} \& \au{Niemela, J.~J.}} \yr{2002}  \at{Quantum
		turbulence}.  \jt{J. Low Temp. Phys.}  \bvol{128},  \pg{167--231}.
	
	\bibitem[Yakhot(2003)]{Yakhot2003}
	{\sc \au{Yakhot, V.}} \yr{2003}  \at{Pressure--velocity correlations and
		scaling exponents in turbulence}.  \jt{J. Fluid Mech.}  \bvol{495},
	\pg{135--143}.
	
	\bibitem[Yui {\em et~al.\/}(2018)Yui, Tsubota \&
	Kobayashi]{QT-coupling-2018-tsu}
	{\sc \au{Yui, S.}, \au{Tsubota, M.} \& \au{Kobayashi, H.}} \yr{2018}
	\at{Three-dimensional coupled dynamics of the two-fluid model in superfluid
		{H}e 4: deformed velocity profile of normal fluid in thermal counterflow}.
	\jt{Phys. Rev. Lett.}  \bvol{120},  \pg{155301}.
	
	\bibitem[Zhou(2021)]{zhou_turbulence_2021}
	{\sc \au{Zhou, Y.}} \yr{2021}  \at{Turbulence theories and statistical closure
		approaches}.  \jt{Physics Reports}  \bvol{935},  \pg{1--117}.
	
\end{thebibliography}


\end{document}